 \documentclass[preprint2]{aastex}

\newcommand{\beq}{\begin{equation}}
\newcommand{\eeq}{\end{equation}}
\newcommand{\bea}{\begin{array}}
\newcommand{\eea}{\end{array}}

\slugcomment{Not to appear in Nonlearned J., 45.}

\shorttitle{Accretion of Heavy Elements } \shortauthors{Zhou et
al.}

\begin{document}

\title{Planetesimal Accretion onto Growing Proto-Gas-Giant Planets}

\author{Ji-Lin Zhou$^1$,  Douglas N.C. Lin$^{2,3}$}
\affil{$^1$Department of Astronomy, Nanjing University, Nanjing
210093, China; zhoujl@nju.edu.cn}

\affil{$^2$UCO/Lick Observatory, University of California, Santa
Cruz, CA 95064, USA; lin@ucolick.org}
 \affil{$^3$KIAA, Peking University,
Beijing 100871, China}

\begin{abstract}

The solar and extra solar gas giants appear to have diverse internal
structure and metallicities. We examine a potential cause for these
dispersions in the context of the conventional sequential accretion
formation scenario, which  assumes the initial formation of cores
from protoplanetary embryos. In principle, gas accretion onto cores
with masses below several times that of the Earth is suppressed by
the energy released from the bombardment of residual planetesimals.
After the cores have attained their isolation masses, additional
mass gain through gas accretion enlarges their feeding zones and
brings a fresh supply of planetesimals. However,  the relatively
low-mass cores have limited influence on exciting the eccentricities
of the newly embraced planetesimals. Due to their aerodynamical and
tidal interaction with the nascent gas disk, planetesimals on
eccentric orbits undergo slow orbital decay. We show that  these
planetesimals generally cannot pass through the  mean motion
resonances of the cores,
 and the suppression of planetesimal bombardment
rate enables the cores to accrete gas with little interruption.
During
 growth from the cores to protoplanets,
the resonance width of protoplanets increases with their  masses.
When the resonances overlap with each other, the trapped planetesimals
become dynamically unstable and their eccentricity excitation is
strongly enhanced. Subsequent gas drag induces the planetesimals
to migrate to the proximity of the protoplanets and collide
with them. This process leads to the resumption and
a surge of planetesimal bombardment
during the advanced stage of the protoplanet growth.  The most
massive intruders are the residual earth-mass protoplanetary
embryos that may survive the passage through the protoplanet
envelopes and increase their core masses.  This mechanism may
account for the diversity of the core-envelope
structure between Jupiter, Saturn and the metallicity dispersion
inferred from the transiting extra solar planets.  During the
final formation stage of the proto-gas-giants, their tidal torque
induces the formation of gaps in the gas disk. This perturbed
structure of gas disk  also leads to the accumulation of planetesimals
outside the feeding zone of the protoplanets. The surface density
enhancement promotes the subsequent buildup of cores for
secondary gas giant planets outside the orbit of the first-born
protoplanets and the formation of eccentric multiple planet systems.
\end{abstract}


\section{Introduction}

Models of the interior structure of giant planets in our Solar System
suggest that giant planets contain heavy elements with total masses up to
several tens Earth masses ($M_\oplus$) (e.g., Wuchterl et al. 2000).
 Due to uncertainties in the equation of state,
the internal distribution of the heavy elements is poorly determined.
Guillot, Gautier \& Hubbard (1997) constructed models of Jupiter which
indicate the presence of a core with mass in the range $0-12
M_\oplus$ and a total mass of heavy elements reaching $11-45
M_\oplus$. With a slight modification in the equation of state used in
these models, Saumon \& Guillot (2004) deduced a smaller upper limit
for the core mass of Jupiter. They find a Jupiter model with core mass of
$0-11M_\oplus$ and total heavy elements of $8-39 M_\oplus$. Their models
suggest Saturn may have a core with mass $9-22M_{\oplus}$ and a total
heavy-element mass of $13-28M_{\oplus}$.

Super-solar metallicity and massive core have also been inferred from
the transit observation of a compact Saturn-mass extra solar planet,
HD149026b (Sato et al. 2005).  However, transit observations of
several other close-in extra solar planets indicate a large dispersion
in the planetary mass-size distribution.  Since the contraction rate
during the  evolution of gas giant planets is determined by the ratios of
their cores to envelope masses (Bodenheimer et al. 2001,
Burrows et al. 2000), a large spread in their mass-radius
relation is indicative of a wide dispersion in their internal
structure.

The existence of massive cores in Saturn and HD149026b provides a
strong support for the conventional sequential accretion scenario
(Perri \& Cameron 1974, Mizuno 1980, Bodenheimer \& Pollack 1986,
Pollack et al. 1996),  which is based on the assumption that gas
giant planets form through three major stages:

\noindent S1) Embryo-growth stage:  protoplanetary cores formed and grew mainly by
the bombardment of planetesimals onto them. Although low-mass cores (up
to a few $M_\oplus$) can also attract a small amount of gas,
the envelopes initially built up at a much slower rate than the
growth of the core. The cores attain isolation masses at the end
of this initial stage.

\noindent S2) Quasi-hydrostatic sedimentation stage:  the
accretion of planetesimals tapers as their supply in the feeding
zone is depleted by the cores. As the dissipation rate also
decreases along with the declining influx of planetesimal
bombardment, thermal energy continues to diffuse out of the
envelope.  This loss of entropy allows a quasi-hydrostatic
sedimentation and the growth of the gaseous envelope.

\noindent S3) Runaway gas-accretion stage: when the mass of the
gas becomes comparable to
that of the core, the rate
of gas sedimentation increases with the intensified flux of radiation
transfer. The characteristic growth time scale of the protoplanets decreases
with their masses.  This runaway stage continues until the gas supply
is exhausted by either the formation of a tidally induced gap near the
protoplanet orbit or the depletion of the entire nascent disk.

While this paradigm has been widely accepted, many uncertainties
remain.  One major issue concerns the protracted transition through
the second stage (S2). In the early models (Pollack et al. 1996),
this stage persists over a time scale longer than the observationally
inferred  depletion time scale of the disk ($\sim$ a few Myr),  because any
increase in the protoplanet  mass also leads to an expansion of its
feeding zone and a resurgence of planetesimal accretion, which tends to
slow down the gas accretion.  A simple extrapolation of the early
models would imply a low probability of gas-giant formation,  which is
incompatible with the observationally inferred ubiquity (with a probability $\sim
0.1-0.15$) of gas giant planets around nearby solar-type stars (Marcy
2005).

Another unsolved issue is the compositional diversity.  Regardless of
the large dispersion in the core mass, the average metallicity of
Jupiter is about twice solar while that of Saturn is an order of
magnitude larger than that of the Sun. There are several potential
mechanisms which may lead to these differences:

\noindent
M1) The critical mass of the cores ($M_{cr}$) needed for the onset of
efficient gas accretion ({\i.e.} from quasi-hydrostatic sedimentation
(S2) to runaway gas-accretion stage (S3)) depends on their
planetesimal bombardment rate ($\dot M_p$) and the rate of radiation
transfer (Ikoma et al. 2000).  Both  processes
may be stochastic and dependent on the inventory of residual
planetesimals and the dust-to-gas ratio.

\noindent
M2) During the runaway gas-accretion stage (S3), a fraction of the pre-existing
cores may be eroded and mixed into the envelope.

\noindent
M3) The metallicity of the accreted gas in the runaway gas-accretion stage (S3)
may depend on the
epoch of protoplanet formation,  since its value increases during the
transition from protostellar to debris disks.

\noindent
M4) During and after the runaway gas-accretion stage (S3),
 gaseous planets may also gain mass in
heavy elements through giant impacts by nearby residual planetesimals
and protoplanetary embryos.  Most intruders are disrupted during their
passage through the envelope.  However, colliding embryos with several
$M_\oplus$ may be able to reach the cores and increase their masses.

Here, we explore how several relevant physical processes may act
together to overcome the growth challenge of gas giants  and introduce
metallicity diversity.  The main aim of this paper is to examine the
dynamical interaction of a growing proto-gas-giant planet with its
neighboring planetesimals and protoplanetary embryos in a gaseous
environment.  This analysis is relevant in two contexts.  Our first
objective is to evaluate whether process M1 can significantly reduce
the planetesimals accretion rate onto relatively low-mass (a few
$M_\oplus$) cores. A reduction of the energy dissipation associated
with planetesimal accretion would enable the gas to settle onto the
cores at a more rapid speed than the early models, thus shorten the
transition from stages of quasi-hydrostatic sedimentation
(S2) to runaway gas-accretion (S3).  The reduced influx of the
planetesimals would also lower the replenishment of dust and the
contribution to the opacity in the envelope. Protoplanetary models
show that the suppression of opacity enhances both the heat flux
through the radiative region and the gas accretion rate (Ikoma,
et al. 2000, Hubickyj et al. 2005).

This promising avenue to bypass the gas-accretion barrier also implies
a limited acquisition of heavy elements during the initial evolution
of proto planets when their mass $M_p$ is only a few Earth masses.  In
order to account for the rich abundance of heavy elements in the gas
giants, especially in their envelopes, we need to consider a possible
mechanism for the resumption of planetesimal-accretion during the
runaway gas-accretion stage (S3) when
the protoplanets have acquired a major fraction of their present-day
gaseous envelopes (with $M_p \sim M_J$ where $M_J$ is Jupiter's mass).
Our second goal is to assess the efficiency of planetesimal accretion
under the influence of process M4. The first objective has
implications for the ubiquity of gas giants whereas the second is
linked to the structural diversity of gas giants.

In \S 2 of this paper, we first present a dynamical model for this
process and estimate the various time scales associated with both
aerodynamical and tidally-induced gas drag. In order to simplify the
problem, we assume a prescribed model for the evolution of protoplanet mass
 which is based on the Bondi formula for idealized,
spherically symmetric, unimpeded accretion. We also utilize an {\it ad
hoc} uniform accretion prescription to illustrate the dominant
physical processes which determine the dynamical evolution of the
system.  In \S3, we show that the combined effects of the
protoplanet's perturbation and the planetesimals' aerodynamical and
tidal interaction with their nascent disks lead to the formation of a
planetesimal gap when $M_p$ is a few $M_\oplus$.  With a reduced rate
of planetesimal bombardment at the onset of the gas accretion, the
envelopes contract more rapidly and the gas accretion time scale can be
shortened from the early models.  In \S4, we show that the planetesimal
accretion rate increases rapidly at the late stage of the
protoplanet's gas accretion.  We also provide evidence to demonstrate
the accumulation of planetesimals outside the asymptotic feeding zone.
This effect can promote the formation of second-generation proto-gas-giant
 planets.  Finally, we summarize our results and discuss their
implications in \S 5.

\section{Dynamical model}

In this section, we study the growth of a protoplanet from the beginning
of quasi-hydrostatic sedimentation stage (S2),
i.e., after its core reaches an isolation mass.
The protoplanet is moving in a gaseous and planetesimal disk.  The
orbits of the protoplanet and planetesimals  are perturbed by the gas drag.
We first estimate the effects of gas drag in \S 2.1. Models of the protoplanet and
planetesimals are given in \S 2.2 and \S 2.3, respectively.

\subsection{Gravitational gas drag}
\label{sec:gasdrag}

There are three physical processes that are acting
on the protoplanet and planetesimals during their evolution:  aerodynamical gas drag,
gravitational tidal drag and dynamical friction.
Dynamical friction on the most massive embryos by the low-mass
planetesimals plays an important role during the stage of runaway
growth of protoplanetary embryos (e.g., Wetherill \& Stewart 1989,
Palmer et al. 1993, Kokubo \& Ida 1996,
 Goldreich et al. 2004). During the subsequent oligarchic stage,
 massive embryos emerge
to perturb the velocity dispersion ($\sigma$) of the residual
planetesimals (Ida \& Makino 1993, Kokubo \& Ida 1998).
 Numerical simulations show that the collisions
generally lead to coagulation and the embryos attain most of the mass
in heavy elements (Kokubo \& Ida 2000, Leinhardt \& Richardson 2005).
In this limit, we assume that the effect of dynamical friction can be
incorporated into the gravitational drag due to embryo-gas
interaction.

\subsubsection{Aerodynamical gas drag}

The gas drag on small particles is in the form of aerodynamical drag
(e.g., Adachi et al. 1976, Tanaka \& Ida 1999).  The
acceleration of a planetesimal with mass $m$ by the aerodynamical drag
has the form
\beq {\bf f}_{\rm aero}=-\frac{1}{2m} C_D \pi S^2 \rho_{g}
|{\bf U}|{\bf U},
\label{faero}
\eeq
where $C_D=0.5$ is the drag coefficient for objects with
large Reynold's number,
%(Landau and Lifthitz 1999),
$S$ is the radius of the planetesimal, $\rho_{\rm g}$ is the density
of gas, ${\bf U}={\bf V}_{\rm k}-{\bf V}_{\rm g}$ is the relative velocity,
${\bf V}_{\rm k}$ and ${\bf V}_{\rm g}$ are the velocity vectors of
the planetesimal's Keplerian motion and gas motion respectively.

The motion of the gas is subject to both the stellar gravity and
its own pressure gradient.  In an unperturbed (by the protoplanet)
region of the disk, the balance of forces in the radial direction
gives
\beq
   \frac{V_{\rm g}^2}{R}=\frac{V_{\rm c}^2}{R}+\frac{1}{\rho_{\rm g}}\frac{dP}{dR},
   \label{vg}
\eeq
where $V_{\rm c} = \sqrt{GM_\ast/R}$ is the pressure-free circular velocity
at the radial distance $R$ to the host star, and $P$ is the gas
pressure. In a stable disk, $P=\rho_{\rm g} c_{\rm s}^2$, where $c_{\rm s}=(\frac{kT}
{\mu m_H})^{1/2}$ is the sound speed of an ideal gas in an isothermal
environment, with $k$ the Boltzmann constant, $m_H$ the proton mass, T the gas temperature
and $\mu$ the average molecular weight of the gas.  We take $\mu=2.34$ for the gas with solar
composition, thus $c_{\rm s}=5.95 \times 10^3 $cm s$^{-1}
\sqrt{T/ {\rm K}}$. Suppose that
\beq
\begin{array}{l}
\rho_g=\rho_{g0} (\frac{R}{\rm 1AU})^{-s_{\rho}}, \\
c_{\rm s}^2=c_{s0}^2 (\frac{R}{\rm 1AU})^{-s_{\rm c}}, \\
T=T_0 (\frac{R}{\rm 1AU})^{-s_T}(\frac{M_*}{M_\odot})^\beta,
\end{array}
 \eeq
 where the subscript $_0$ denotes the  value of the corresponding
 quantity at 1AU (and $M_*=M_\odot$ in the equation of $T$).
 From equation (\ref{vg}) we get
 \beq
 V_{\rm g}= V_c  (1-2 \eta (R))^{1/2},
 \label{vgeta}
 \eeq
 where
\beq
\eta= \frac{(s_{\rho}+s_{\rm c})}{2} (\frac{c_{\rm s}}{V_{\rm c}})^2,
\label{eta}
\eeq
and
\beq
\left( \frac{c_{\rm s}}{V_c} \right)^2=4 \times 10^{-6}
\left(\frac{T_0}{K} \right) \left(\frac{R}{\rm 1AU} \right)^{1-s_T}
  \left(\frac{M_*}{M_\odot} \right)^{\beta-1}.
  \label{csovk}
\eeq

%where $V_{\rm k} = \Omega_{\rm k} R$ and $\Omega_{\rm k} = \sqrt{G M_\ast/R^3}$
%and the Keplerian velocity and angular frequency at the radial
%distance $R$ to the spin axis of disk,
 %\beq
 %\eta \equiv = - {1 \over 2 \rho_{gas} R \Omega_{\rm k}^2} {\partial p
 %\over \partial R},
 %\eeq

Throughout this paper, we use the minimum mass nebula model
(Hayashi et al. 1985) to evaluate fiducial model parameters, though our
basic algorithm can be applied to a more general disk.  In this
model, the gas is heated to an equilibrium temperature by the central
star with temperature
 \beq
 T =  280\left( {R \over 1 {\rm AU}} \right)^{-1/2} \left(
 {M_\ast \over M_\odot} \right).
 \eeq
  The surface density of the gas disk is given by
 \beq
 \Sigma_{\rm g}= f_g \Sigma_{g0}  \left(\frac{R}{1{\rm
 AU}} \right)^{-3/2},
 \label{mingas}
 \eeq
where $\Sigma_{g0}=1700 ~{\rm g ~cm^{-2}}$, $f_g$ is
a scaling factor so that $f_g=1$ corresponds to the minimum mass nebula.
The corresponding density of gas disk is
 \beq \rho_{\rm g}= 1.4 \times 10^{-9} {\rm g ~cm^{-3}} f_g
 \left(\frac{R}{1{\rm AU}} \right)^{-11/4}.
 \label{rhogas}
 \eeq
 Since we have $s_\rho=11/4,s_c=1/2$, substituting them into  equations
  (\ref{eta}) and (\ref{csovk}),
 $\eta$ has the form of
 \beq
 \eta(R)=0.0018 \left(\frac{R}{1AU} \right)^{1/2}.
 \eeq
In more realistic models, the magnitude of $\eta$ can be modified by the surface irradiation,
internal viscous dissipation, and the radiation transfer through the
disk (Garaud \& Lin 2007).

Assuming the protoplanet or a planetesimal has a density $\rho$,
 its radius can be expressed as
 \beq
 S=5.2 \times 10^{-3} {\rm AU} \left(\frac{M}{M_{\odot}}
 \right)^{1/3} \left(\frac{\rm 1 ~g~cm^{-3}}{\rho} \right)^{1/3}.
 \label{S}
 \eeq
In terms of the above expression, equation (\ref{faero}) can be expressed as
 \beq
 \begin{array}{ll}
 {\bf f}_{\rm aero}
 = & - 10^{-7}C_D f_g~{\rm AU}^{-1}
 \left(\frac{R}{1{\rm AU}} \right)^{-11/4}  \\ &
   \left(\frac{M}{M_\odot}
\right)^{-1/3} \left(\frac{\rho}{\rm 1~ g ~cm^{-3}}\right)^{-2/3}
  |{\bf U}|{\bf U}.  \end{array}
  \label{faofin}
\eeq
%where $f_g = \Sigma_{g}/\Sigma_{g0}=\rho_{\rm g}/\rho_{\rm g0}$ is a scaling factor.

The aerodynamical gas drag decreases the semi-major axis $a$, eccentricity
$e$, and inclination $i$ of the planetesimal orbit. The average
time scales for the evolution of these orbital elements are given by
Adachi et al. (1976)
 \beq
 \begin{array}{ll}
 \frac{1}{\tau_{a,a}} & \equiv  \frac{1}{a} \left(\frac{da}{dt}\right)_{\rm aero}
  \\  & =  -\frac{2}{\tau_{\rm aero}} \left(\frac58 e^2+\frac12 i^2+ \eta^2
 \right)^{1/2}    \left( \eta+\frac{17}{16}e^2+\frac18 i^2 \right)
 \\
 \frac{1}{\tau_{a,e}} & \equiv  \frac{1}{e} \left(\frac{de}{dt} \right)_{\rm aero}
 =\frac{2}{i} \left(\frac{di}{dt} \right)_{\rm aero}
 \\ & =
 -\frac{1}{\tau_{aero}} \left(\frac58 e^2 +\frac12 i^2+ \eta^2 \right)^{1/2},
 \label{daaero}
 \end{array}
 \eeq
where $\tau_{\rm aero}$ is a time scale given as:
 \beq   \begin{array}{ll}
 \tau_{\rm aero}& =\frac{2m}{\pi C_D s^2 \rho_{\rm g} V_{\rm k}(a)} \\
    & \approx
 \frac{3.5 \times 10^3 {\rm yr}}{f_{\rm g}} \left(\frac{M}{10^{17} {\rm g}} \right)^{1/3}
 \left(\frac{\rho}{\rm 2g cm^{-3}} \right)^{2/3}
 \left(\frac{a}{\rm 5AU} \right)^{13/4}.
 \end{array}
 \label{tgas}
 \eeq
Note that $\tau_{a,a} \sim \tau_{a,e} / (\eta+ e^2+i^2) > > \tau_{a,e}$.

\subsubsection{Tidal effect of gas disk}

Tidal interaction between  a gas disk and a
protoplanet  leads to an effect similar to gas drag,
which is particularly important for the dynamical
evolution of protoplanets with large masses (Goldreich
\& Tremaine 1980, Ward 1986, 1989, 1997, Artymowicz 1993).  We adopt
the form of acceleration from Kominami and Ida (2002):
 \beq
 \mathbf{f}_{\rm tidal}=-\frac{{\bf V}_k-{\bf V}_{\rm g}}{\tau_{\rm tidal}},
 \label{tidal}
 \eeq
where ${\bf V}_{\rm g}$ is the velocity of gas motion,
and we consider it in circular orbits, and $\tau_{\rm tidal}$ is the time scale defined as
(Ward 1989, Artymowicz 1993)
 \beq
 \begin{array}{ll}
 \tau_{\rm tidal} & =\left(\frac{M}{M_\ast} \right)^{-1}
 \left(\frac{\Sigma_{g}a^2}{M_\ast} \right)^{-1}
 \left(\frac{c_{\rm s}}{V_{\rm c}} \right)^4 \Omega_{\rm k}^{-1} \\
&  \approx {5 \times 10^{4} {\rm yr} \over f_g}
 ~ \left(\frac{M}{10^{27} {\rm g}} \right)^{-1} \left(\frac{a}{5 {\rm AU}} \right)^2,
\eea
 \label{ttidal}
 \eeq
where $\Omega_k$ is the Keplerian frequency of circular motion.
 Since dynamic friction  of planetesimals
 on protoplanets have similar expressions of acceleration and
 time scale as those of the disk tide, i.e., equations
 (\ref{tidal}) and (\ref{ttidal}) (see Appendix of Kominami \& Ida 2002),  we
 do not consider the effect of dynamical friction particularly in this work.

Tidal drag decreases the eccentricities  and
inclinations of the embryo orbits.
In principle, tidal interaction between embryos and the
disk also lead directly to the decay of embryo orbits (Ward 1997). The
rate of this ``type I migration'' is determined by an imbalance in
the torque from disk regions interior and exterior to the embryo
orbits.  Linear analysis for this process is evaluated for
idealized  background surface density (Tanaka
 et al. 2002). The results of linear analysis imply that,
in a solar nebula environment, embryos more massive than the Earth
would migrate rapidly toward their host stars, thus  gas giants
would rarely form (Ida \& Lin 2007).  In general, both intrinsic (Rice
\& Armitage 2003, Laughlin et al. 2004,  Nelson \&
Papaloizu 2003) and self-excited turbulence (Koller et al. 2003) may
lead to nonlinear evolution of disk structure, readjustment of the
torque distribution, and the suppression of type I migration.
However, these structural adjustments do not modify the
efficiency of eccentricity damping since the contribution from both sides
of the disk is cumulative rather than cancelling.  Thus, we include
here the effect of tidally-induced eccentricity damping but neglect
that of type I migration.  Nevertheless, there is an associated change
in the semi-major axes of the planetesimals and embryos.  Within
$O(e^2,i^2)$, the average time scales for the evolution of these
orbital elements are given by,
 \beq
 \begin{array}{l}
 \frac{1}{\tau_{t,a}} \equiv \frac{1}{a} \left(\frac{da}{dt} \right)_{\rm tidal}
 =-\frac{1}{8\tau_{\rm tidal}}( 5e^2 + 2i^2 ) \\
  \frac{1}{\tau_{t,e}} \equiv \frac{1}{e} \left(\frac{de}{dt} \right)_{\rm tidal}
 =-\frac{1}{\tau_{\rm tidal}} \left(1-\frac{13}{32} e^2 -\frac18 i^2 \right) \\
 \frac{1}{i} \left(\frac{di}{dt}\right)_{\rm tidal}=-\frac{1}{2\tau_{\rm tidal}}
 \left(1+\frac{11}{16}e^2 +\frac{3}{16} i^2 \right).
 \end{array}
 \label{datide}
 \eeq
See Appendix for a brief derivation.

\begin{figure}
\vspace{4cm}
\includegraphics{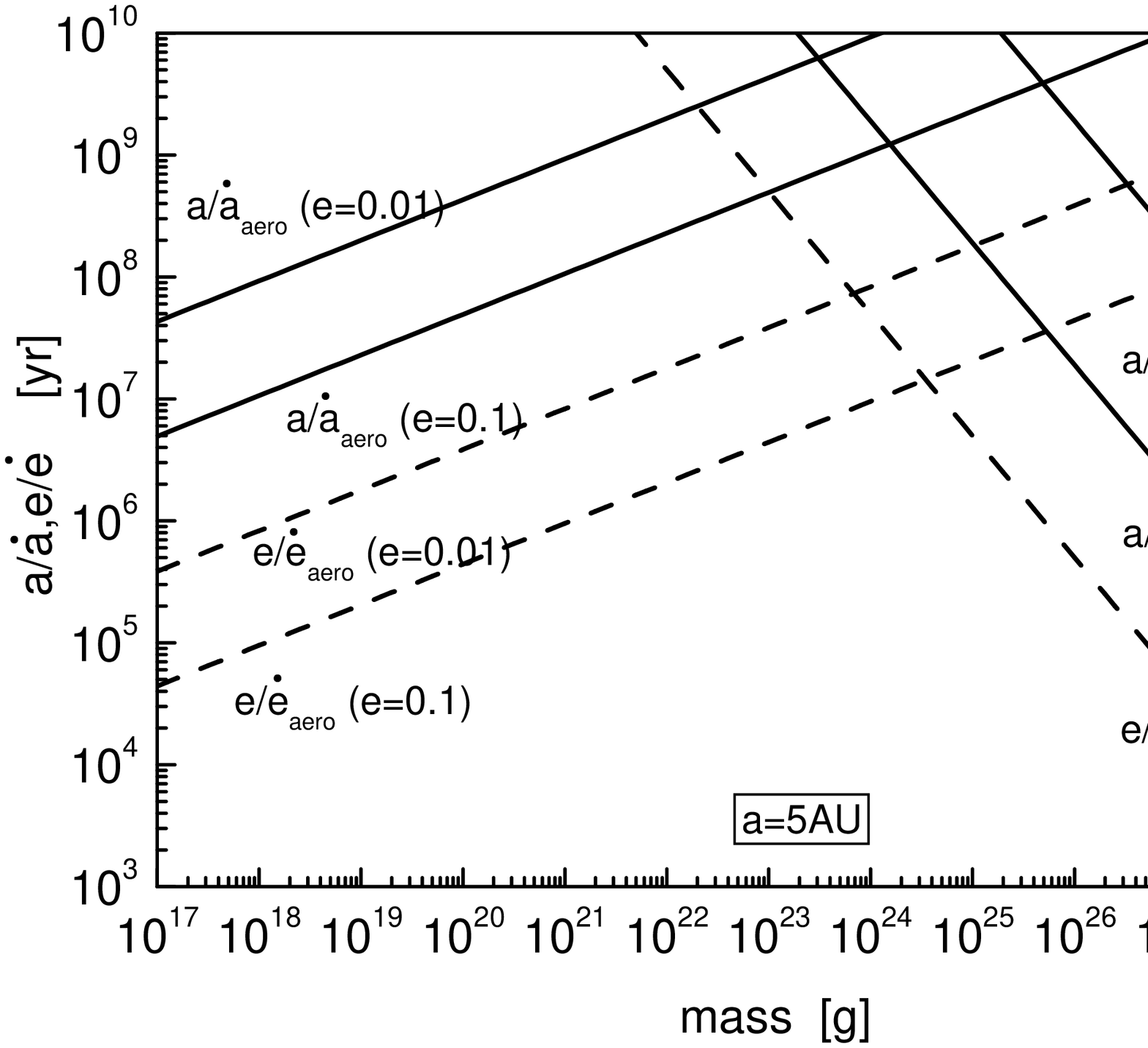}
 %\plotone{f1.eps}
 \caption{\small Damping time scales in the
semi-major axis $a$ and eccentricity $e$  due to the aerodynamical
(Eq. [13]) or tidal drag (Eq.[17]) for different mass of a
planetesimal (or a protoplanet) located at 5 AU. For aerodynamical
drag, a density of  $\rho=2 {\rm ~g ~cm^{-3}}$ is assumed.
 \label{fig1}}
\end{figure}

The times scales for the orbital decay of planetesimals or protoplanets under aerodynamical
and  tidal drag are shown in Fig.1 (See page 9).
  During the disk evolution
and depletion (on a time scale $\sim 10^{6-7}$ yr), aerodynamical
drag is more effective for small planetesimals with mass $\le 10^{23}$g,
 while tidal drag is more important for embryos with mass
 $ \geq 10^{24}$g.
The surface density of the gas is globally depleted on a time scale
$\tau_{\rm dep} \sim 1-3 {\rm Myr}$, so the magnitude of all
time scales of the gas drag increases.

\subsection{Protoplanet model}

In this paper, we study the growth of an isolated protoplanet as the
progenitor of Jupiter.  We analyze the dynamical evolution of its
nearby residual planetesimals and embryos subject to the
perturbation of the protoplanet and gas drag.
 We start at the beginning of the quasi-hydrostatic
sedimentation stage (S2) of the protoplanet formation.
 The following simplifications are adopted in our
simulations:

\noindent
i) After its core has acquired an isolation mass, the protoplanet is
assumed to accrete gas with prescribed rates. The gas accretion model
is described in \S 2.2.1.

\noindent
ii) The dynamical feedback of planetesimals to the protoplanet is
neglected because random orbital phases usually counteract each other.
The accumulative feed-back perturbations by its co-existing
nearby embryos on the protoplanet will be studied in
the future.

\noindent
iii) The protoplanet does not have a significant radial excursions
during the evolution.

The last approximation is consistent with
neglecting type I migration.  Although close-in planets may have
attained their present-day orbits through extensive type II migration
(Lin et al. 1996), these events occur after the
proto-gas-giant planets have already acquired most of their masses and
on the viscous evolution time scale of the protostellar disks (Lin \&
Papalozou 1986).  Here we focus our investigation primarily on the
acquisition of planetesimals during the  formation stages of a protoplanet.

Under the above simplifications, we place a protoplanet in an orbit
with elements $a_p=5$ AU, $e_p=0.01$, and $i_p=0.005$. According to
equation (\ref{datide}), the eccentricity and inclination of such an isolated
protoplanet embedded within the solar nebula would be damped by the
tidal drag soon, on a time scale shorter than its growth time scale prior to
gap formation but longer than the synodic period of nearby embryos and
planetesimals (see below). In dynamical equilibrium, the recoil
motion of the protoplanet is balanced by the tidal damping to yield
the assumed eccentricity and inclination, especially during the onset
of the quasi-hydrostatic sedimentation stage (S2) when the mass of
the protoplanet is modest.
 During the runaway gas-accretion stage (S3), a gas gap forms near the protoplanet and
$\tau_{\rm tidal}$ becomes longer than its growth time through gas accretion.

\subsubsection{Inner and outer feeding zones}
\label{sec:feeding}

Planetesimals grow into protoplanetary embryos (which significantly
perturb the motion of their neighboring planetesimals) and cores
(which accrete gas) through cohesive collisions (Safronov 1969).  The
region from where a protoplanet has a non-zero probability to accrete
planetesimals during a single azimuthal passage is referred
to as its feeding zone. Neglecting the host stars' tidal perturbation
on their equation of motion (but including the Keplerian
shear), all planetesimals with semi-major axis $a$ and eccentricity $e
> \delta_a \equiv \vert a/a_{\rm p} -1 \vert$ are contained in the
 feeding zone of the protoplanet centered on its semi-major axis $a_{\rm p}$.
However, many synodic periods ($\tau_{\rm syno} \sim 2 P_{\rm k} / (3\delta_a)$,
where $P_{\rm k}$ is the period of Kepler motion) may be needed before
close encounters can occur.  The accretion rate onto the protoplanet
with a mass $M_{\rm p}$ and radius $R_c$ is (Safronov 1969)
 \beq
 \bea{ll}
 \dot M_{\rm p} & \sim \pi R_c^2 \rho_{\rm d} \left( {2 G M_{\rm p} \over R_c \sigma^2} \right)
 \sigma \\
 & \sim \left( {2 \pi \over e^2 + 2.25 \delta_a^2} \right)
 \left( {R_c \over a_{\rm p}} \right)
 \left( {M_{\rm p} \over M_\ast} \right) \Sigma_{\rm d} a_{\rm p}^2 \Omega_{\rm k},
 \eea
 \label{eq:mdotp}
 \eeq
where $\sigma = \sqrt{e^2 + 9 \delta_a^2/4 } \Omega_{\rm k} a_{\rm p}$, $H_{\rm p} \sim
\sigma/\Omega_{\rm k}$, $\rho_{\rm d} \sim \Sigma_{\rm d}/ 2 H_{\rm p}$, and $\Sigma_{\rm d}$  are the
velocity dispersion, scale height, spatial and surface density of the
planetesimals, respectively.  The eccentricities of the planetesimals are excited by
the secular perturbation of the protoplanet,   and the average excursions of
the eccentricities and inclinations of the planetesimals per synodic encounter can be expressed
as (Hasegawa \& Nakazawa 1990),
 \beq
<\Delta e>\approx  \left(\frac{1.9h}{\delta_a} \right)^3 \delta_a,
~~  <\Delta i>\approx
\left(\frac{1.3h}{\delta_a} \right)^3i,
 \label{de}
 \eeq
where $h \equiv (M_{\rm p} / 3M_\ast)^{1/3}$ is the scaled
Hill's radius of the protoplanet.  Note that secular perturbation does not significantly
change the semi-major axes of  the planetesimals.  Over $\tau_{\rm syno}$ the eccentricities
of planetesimals with mass larger than $10^{18}$ g are not
significantly damped by either aerodynamical or tidal gas drag.  In
principle, nonlinear diffusion can lead to further eccentricity growth
in a gas-free environment.  But it proceeds on time scales much longer
than $\tau_{\rm syno}$ and is effectively suppressed by the gas damping
effect (Zhou et al.  2007).

Equation (\ref{de}) implies $e \sim \delta_a$ at $\delta_a \sim 1.5 h $.
 We define this location to be the boundary between the inner and outer
feeding zone.  In the inner feeding zone, the planetesimals undergo
radial excursions which cross the protoplanet orbit at each
azimuthal conjunction. The two-body formula is a reasonable
approximation to their encounters.  In the outer feeding zone where
$\delta_a > 1.5 h$, two-body effects alone cannot lead to close
encounters.  However, when the tidal perturbation of the host star is
also included in the equation of motion, the feeding zone expands to
include more distant planetesimals. A simple approach to approximate
the  orbits of the planetesimals is to use a restricted three-body
approximation.  Although the perturbation of the protoplanet induces
periodic (or chaotic if the planetesimal has relatively large energy)
variations on the planetesimals' star-centric orbital elements
$a,e,i$, the motions of the planetesimals are constrained by the
Jacobi integral, which can be expressed as
 \beq
 E_J= \frac{1}{2}(e^2+i^2)-\frac{3 }{8}\delta_a^2 +\frac{9}{2}h^2+O(h^3).
 \label{EJ}
 \eeq

Planetesimals with positive Jacobi energy reside in the feeding zone
(Hayashi et al. 1977), i.e. they have finite close encounter
probability per $\tau_{\rm syno}$.  In the $(a,e)$ and $(a,i)$ planes, the
boundary of the feeding zone $E_J=0$ is on a branch of hyperbolic
curves. The half width of the feeding zone is $\delta_a \sim 2\sqrt{3} h$
for planetesimals with $e < \delta_a$. In the outer feeding zone
where $\delta_a/h \sim 1.5-3.5$, $e < \delta_a $,  the planetesimals can engage in
close encounters and possible physical collision with the protoplanet only within
certain ranges of the longitude of periapse.
Through nonlinear diffusion, the orbit-orientation angle undergoes random
walk (Zhou et al. 2007) and over many $\tau_{\rm syno}$'s
planetesimals may occasional venture into the
collision ``key holes'' of the protoplanet.  The range of the phase for direct
collision decreases with $\delta_a$ and vanishes in the limit
$\delta_a > 2\sqrt{3} h$.  Thus, the  accretion rate of
planetesimals onto the  protoplanet from the outer feeding zone with $\delta_a \sim
1.5-3.5 h$ is significantly reduced from the value of $ \dot M_{\rm p}$ in
equation (\ref{eq:mdotp}).

However, in a gaseous environment, the accretion of planetesimals
in the entire feeding zone can be enhanced by the gas drag effect,
especially in the outer feeding zone.  Although the eccentricity
excitation of the planetesimals in this location is limited by equation
(\ref{de}), tidal damping of their eccentricities leads to orbital decay. In
\S\ref{sec:pgap} we show that planetesimals drift in from the
outer edge of the feeding zone to the proximity of the protoplanet
where the gravitational perturbation is intense and close
encounters occur more frequently. It is such induced orbital decay
rate rather than the protoplanet accretion rate ($ \dot M_{\rm p}$)
that determines the time scale for the clearing of the feeding zone.
 Along the way, these
migrating planetesimals pass through the  mean motion
resonances of the protoplanet and their eccentricities
are excited to large amplitudes. The
subsequent gas drag leads to orbital decay. With relative small
$\tau_{\rm aero}$ and $\tau_{\rm tidal}$, small planetesimals and
large embryos pass through the mean motion resonances.  But some
intermediate-mass planetesimals have relatively long $\tau_{t,a}$
and $\tau_{a,a}$ and may be trapped by the mean motion
resonances (see \S\ref{sec:resonance}).

\subsubsection{Isolation core mass and gas accretion model}
\label{sec:isolat}

A planetary core temporarily halts its growth by accreting planetesimals
when it attains an isolation mass $M_{\rm iso}$ (Lissauer 1987).  If
all the planetesimals in the feeding zone can be accreted by the
protoplanet, the magnitude of $M_{\rm iso}$ would be determined by
$\Sigma_{\rm d}$ and the width of the core feeding zone ($\Delta a$):
 \beq
  M_{\rm iso}=2\pi \Sigma_{\rm d} a_{\rm p}\Delta a.
 \eeq
It is useful to scale the magnitude of $\Sigma_{\rm d}$ with that of the
fiducial minimum solar nebula (Hayashi 1981) outside the snow line,
\beq
\Sigma_{\rm d, min} = 30 (a/ 1{\rm AU})^{-3/2} ~{\rm g ~cm^{-2}},
\label{dust}
\eeq
 by a multiplicative factor $f_{\rm d}$.  Numerical simulations (Kokubo \& Ida
2002) indicate that the feeding zone of a core has a width of $\Delta_a =
10 a_{\rm p} h$ which is slightly larger than twice $2 \sqrt{3} a_{\rm p} h$ for
low-eccentricity planetesimals. This minor expansion is
due to the eccentricity excitation associated with nonlinear
diffusion (rather than linear secular perturbation) over the time
scale for reaching the feeding zone, $\sim 3 M_{\rm iso}/\dot M_{\rm p}
(M_{\rm iso})$.  From these values, we obtain
 \beq
 M_{\rm iso}=2 M_{\oplus} f_{\rm d} ^{3/2}
 \left(\frac{a_{\rm p}}{5{\rm AU}} \right)^{3/4}.
 \eeq
We place our protoplanet at the present-day location of Jupiter,
i.e. at 5AU which is slightly outside the snow line. In a minimum
mass nebula, this radius is a preferred location for the onset of gas
giant formation because 1) the isolation mass of the embryos is a few
$M_\oplus$ and 2) the time scale for the buildup of the embryo with
isolation mass is comparable to or shorter than the disk depletion
time scale (Ida \& Lin 2004).  We take $f_{\rm d}=2$ in this study, so
the protoplanet core in $5$ AU has a mass of $M_{p0}=5.67~M_{\oplus}$
before the onset of gas accretion.

Cores with isolation masses attract nearby gas because their surface
escape speed is much larger than the sound speed of the disk
gas. Nevertheless, heat is released from the contraction of gas onto
the cores.  When the protoplanet  mass is low (a few $M_\oplus$),
inefficient heat transfer in its envelope leads to the buildup of a
high pressure gradient to balance its gravity and slow down the
accretion rate in the quasi-hydrostatic sedimentation stage (S2).
 Early numerical models of protoplanetary
structure indicate that the main heat-diffusion bottleneck is in the
outer radiative region (Pollack et al. 1996). The gas accretion
rate would be greatly enhanced if the grain-dominated opacity is
suppressed (Ikoma et al. 2000, Hubickyj et al. 2005).
  A possible
mechanism for major opacity reduction is  dust coagulation and
sedimentation through the outer radiative region.
However, the grains may also be replenished by much larger colliding
planetesimals which disintegrate during their passages through the
envelope.  The clearing of planetesimals from the feeding zones of the
accreting cores would greatly reduce the resupply rate.

As a first approximation, we assume the heat transfer barrier can
be bypassed by the depletion of the feeding zone,  and the process of
gas accretion is not impeded by the radiative feedback.  At 5AU, when
the protoplanet mass is well below $M_J$, its Bondi radius $R_b =G M_{\rm p}
/c_{\rm s}^2$ is smaller than its Hill's radius $R_h = h a_{\rm p}$ and the disk scale
height $H/a_{\rm p}=  c_{\rm s}/ V_{\rm k}= 0.07(a_{\rm p}/5~{\rm AU})^{1/4}$ in a minimum
mass nebula.  For such a protoplanet, the disk gas in the background
is homogeneous and the tidal effect of its host star can be neglected.
For most of our calculations, we adopt the conventional Bondi formula
(Frank et al. 2002) for spherical accretion, in which the
accretion rate is given as
 \beq \dot{M}_{BD} =
 \frac{ \pi G^2 \rho_{\rm g}}{c_{\rm s}^3}\bar{\alpha} M^2 \equiv \xi
 \bar{\alpha} M^2, \label{bd}
 \eeq
where $\bar{\alpha}$ is a constant  of order unity determined by the state equation
of the gas.
For a protoplanet  at 5 AU, we find $ \xi \approx 1.2\times 10^{-32}f_{\rm g} {\rm g}^{-1}
{\rm yr}^{-1} =24 f_{\rm g} M_{\small \odot}^{-1} {\rm yr}^{-1}$.  By integrating
equation (\ref{bd}), we obtain
 $M=M_0 [1-(1-M_0/M_f)(t-t_0)/\tau_{\rm grow}]^{-1}$ for $t-t_0 \le \tau_{\rm grow}$,
  where $t_0$ is the epoch when the accretion begins,
$\tau_{grow}$ is the time scale of the protoplanet growth to a mass of $M_{\rm f}$,
 \beq
 \tau_{\rm grow}\equiv \frac{1}{\bar{\alpha}
 \xi} \left(\frac{1}{M_0}-\frac{1}{M_{\rm f}} \right)\approx \frac{1}{24
 f_{\rm g} \bar{\alpha}\mu_0} {\rm yr}, \label{tgt}
 \eeq
where $\mu_0=M_0/M_{\odot}$, $M_0$ is the mass of protoplanet at
the onset of gas accretion, and we assume that $M_{\rm f} >> M_0$ in the
approximation of equation (\ref{tgt}).  In a minimum mass nebula, a
$5.67~M_{\oplus}$ protoplanet at 5 AU has $\tau_{\rm grow}
\sim 10^5$ yr.  This time scale only refers to the growth time scale
associated with the gas accretion.  The emergence of such a massive core
and the transition from stages of embryo-growth (S1) to  quasi-hydrostatic sedimentation
(S2) may take longer time.

The validity of the assumed homogeneous, unimpeded, spherical accretion
flow onto the protoplanet is questionable when its mass becomes comparable
to that of Jupiter because $R_b \sim R_h \sim H$.  The effect of
differential rotation in the disk and the tidal torque by the host
star channel the accretion flow through a protoplanetary disk.  We are
primarily interested in the dynamical evolution of residual
planetesimals and embryos near the outer regions of the feeding zone
($\delta_a > h$) which is not strongly perturbed by the distribution
of gas within $R_h$.  As the protoplanet attains its asymptotic mass,
its tidal interaction with the gas disk  leads to the formation of a gap
(see \S{\ref{sec:gasgap}).  The open of gap reduce the accretion rate
from the unimpeded $\dot M \sim 10^{-3} M_J$ yr$^{-1}$  in
equation (\ref{bd}) by several orders of magnitude
(Dobbs-Dixon et al. 2007).  However, this transition occurs rapidly and we can approximate
it with an abrupt termination of its growth.  In order to take into
account the uncertainties in the boundary conditions, we consider a
range of values for $\bar{\alpha}$  and  $f_{\rm g}$ so that
 $\tau_{\rm grow}=10^{3}-10^6$ yr in the following calculations.

In the Bondi model, the growth time scale decreases with the mass of the
protoplanet. As we show below, this growth pattern can lead
to the initial quenching of planetesimal bombardment and the
late-stage capture of residual embryos and planetesimals.  In order to
highlight this behavior, we consider a second series of models
with an {\it ad hoc} prescription in which a constant gas accretion
rate,
 \beq
\dot{M}_{LN}=\frac{1}{\tau_{\rm grow}}(M_{\rm f}-M_0),
 \label{ln}
 \eeq
onto the protoplanet is assumed.  Figure 2 shows the evolution of
protoplanetary mass by accreting gas on a time scale of $\tau_{\rm
grow}=10^5$~yr according to these two models.
The initial mass of the isolated core is set as
$5.67 ~ M_{\oplus}$.  Gas accretion is most effective around
$\tau_{\rm grow}=10^5$~yr in the Bondi model.

\begin{figure}
%\plotone{f2.eps}
 \vspace{5cm}
\includegraphics{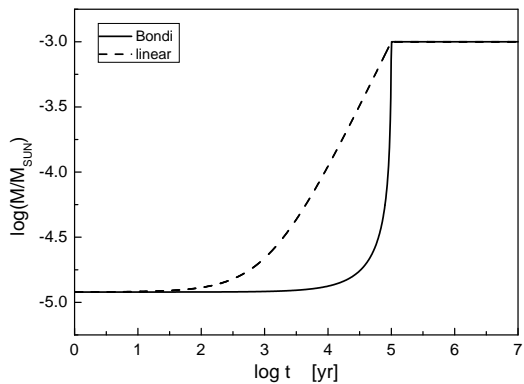}
\caption{\small Increase of protoplanet  mass in  two gas-accretion models:
Bondi accretion (Eq. [\ref{bd}])
 and linear accretion  (Eq. [\ref{ln}]) on a time scale $\tau_{\rm grow}=10^5$ yr.
The initial mass of the planet is $5.67~ M_{\oplus}$ and the final
mass is 1 Jupiter mass.
 \label{fig2}}
\end{figure}

\subsubsection{Gap opening in gaseous disk}
\label{sec:gasgap}

When a protoplanet grows to a sufficiently large mass, a gap in the gas
disk forms around its orbit (Lin \& Papaloizou 1979). The formation
of the gap has two important effects.  As we have already indicated
above, gap formation greatly reduces the accretion rate onto the
protoplanet and effectively terminates its growth.  The clearing of
the gas also reduces the magnitude of the drag on the residual
planetesimals and embryos (see \S\ref{sec:gasdrag}). Here we briefly
describe our prescription for the emergence of tidally induced gaps in
the disk.

The critical mass of a planet ($M_c$) over which it can open a gap is
determined (Lin \& Papaloizou 1993) by a viscous and a thermal
condition,
  \begin{equation} \frac{M_{c, v}}{M_{\odot}}=\frac{40\nu}{a_{\rm p}^2\Omega_{\rm k}}= 40\alpha
  \left(\frac{c_{\rm s}}{a_{\rm p}\Omega_{\rm k}} \right)^{2},
  \end{equation}
  \begin{equation} \frac{M_{c, t}}{M_{\odot}}=3
  \left(\frac{H}{a_{\rm p}}\right)^3,
  \end{equation}
where $\nu= \alpha c_{\rm s}^2/\Omega_{\rm k} $ is the kinematic viscosity, and
$\alpha$ is a parameter (Shakura \& Sunyaev 1973). At 5AU in a minimum
mass nebula, a protoplanet continues to grow until its mass
reaches $M_{c,t} \sim 300~ M_{\oplus}$ so that both the
thermal and viscous conditions are satisfied .

The minimum width of the gap is the  Hill's radius $R_h$
of the protoplanet.
Numerical simulation indicates that the gap extends to $2-3 R_h$. As
the mass of the protoplanet mass approaches that of Jupiter,
its unimpeded growth time scale reduces to less than $10^3$ yr,
 which is shorter than the
viscous diffusion time scale across all regions of the disk wider than
$R_h$.  Consequently, gas within a gap-feeding zone is rapidly depleted.
In the low-viscosity limit (where $\alpha < < 1$), we estimate the
half-width of the gap $\Delta$ on the assumption that all the gas in
the gap is accreted on to the protoplanet and contributes to its
asymptotic mass. Let us denote the mass increment as $\delta M$.  In the absence of viscous
mass diffusion, $\delta M= 8 \pi\int_{a_{\rm p}}^{a_{\rm p}+\Delta}\rho_{\rm g} a H
da $, and the half-width of the gap is given by
 \beq \frac{\Delta}{a_{\rm p}}=160 \left(\frac{\delta M }{M_{\odot}}
  \right) \left(\frac{a_{\rm p}}{5AU} \right)^{-1/2}.
  \label{gapwidth}
 \eeq
From the above minimum-mass nebula parameters and the present-day
$a_{\rm p}=5AU$, $\delta M+M_{\rm p0} =10^{-3}M_{\odot}$, we find
$\Delta \approx 0.8 $ AU, which is approximately $2.3 R_h$ and
comparable to the value found in numerical simulations (cf Bryden
et al. 2000).

%It may also be modified in the vicinity of the protoplanet by its
%tidal torque (see \S\ref{sec:gasgap}).
% When a sufficiently
% massive protoplanet satisfied both the viscous and thermal
% criteria, a gap is formed near its orbit to terminate its gas
% accretion (see \S\ref{sec:gasgap}).

The gap-opening  in the gas disk near a massive protoplanet may change
the kinematics of nearby planetesimals.
 At the inner edge of the gas
gap, $\eta$ is enlarged due to the sharp pressure contrast,  and the
orbital decay rate of planetesimals is enhanced.  However, at the
outer edge of the gas gap, the direction of gas pressure is
inwards and the pressure gradient is
positive, resulting in $V_g> V_c$ according to equation (\ref{vg}), and
$\eta$ is negative by equation (\ref{vgeta}) (Lin \&
Papaloizou 1993, Bryden et al. 2000). Thus the tail wind of
the gas induces the planetesimals to migrate outward.  The severe
depletion of gas in the gap region also reduces the drag at that
location.  However, these feedback effects are negligible until
the protoplanet has opened a gap and attained most of its
asymptotic mass. After the gap formation,
the sign of $\dot a$
becomes positive (see Eq. [\ref{daaero}]) for low-eccentricity planetesimals (with $e^2 <
\vert \eta \vert$) and they move beyond the gap,  whereas the
high-eccentricity planetesimals continue to migrate inward.

The modification of $V_{\rm g}$ near the outer edge of the gas gap
also locally changes the sign of the tidal torque on the protoplanetary
embryos (see Eq. [\ref{tidal}]).  In principle, this process
induces the embryos to undergo outward migration, analogous to the
type I inward migration. But similar to the aerodynamical drag, the
damping of the embryo eccentricities dominates the
evolution of their semi-major axes.

 As the depletion of gas in the gap reduces the local density
of the gas disk, we take ${\bf f}=0$  in
the corresponding aerodynamical and tidal drag formulas (Eqs.
[\ref{faero}] and [\ref{tidal}]) inside the gap in our simulations.
We assume that gas in the gap is depleted on a time scale $\tau=\Delta
r^2/\nu\sim H^2/\nu $, which gives $\tau \approx 2 \alpha^{-1}$
yr.  Thus for $\alpha=10^{-3}$, we have $\tau\approx 2000$
yr.  However, to simplify the problem, we neglect the migration effect
due to  gap formation, which is very interesting and
will be studied elsewhere.

\subsection{Planetesimal model}

In this paper, we study the evolution of planetesimals with initial
semi-major axes $a\in [3.2,8.5]$ AU, i.e., $a\in [a_{\rm p}-5a_{\rm p} h_{\rm f},
a_{\rm p}+10a_{\rm p} h_{\rm f}]$ with $a_{\rm p}=5$AU, $h_{\rm f}=(\mu_{\rm f}/3)^{1/3}$ and
$\mu_{\rm f}=10^{-3}$. We choose a larger region for the outer part, because
due to gas drag and eccentricity damping, planetesimals on both side
of the protoplanet suffer orbital decay.  Exterior to the protoplanet,
planetesimals migrate toward its feeding zone.

Numerical simulations indicate that during the oligarchic growth, the
feeding zone of the most massive embryos may indeed contain less
massive planetesimals and embryos (Kokubo \& Ida 2002).  In order to
examine the mass-dependent collision probability of planetesimals with
a growing protoplanet, we carry out several simulations,
each with a population of uniform mass planetesimals.  The individual
planetesimal mass under investigation ranges from $10^{17}
-10^{27}$g. The planetesimals are treated as test particles in the
simulation (i.e. they do not impose any gravitational
perturbation on each other or on the protoplanet) except when we
evaluate the magnitude of the gas drag. The low-mass range corresponds
to km-size objects,  which can withstand the aerodynamical drag in a
minimum mass nebula with $\tau_{a, a} > \tau_{\rm dep}$ and be retained by
the disk.  We did not simulate the interaction between a protoplanet
and embryos more massive than $10^{27}$g because it would not be
adequately approximated by the restricted three-body approach.  A more
comprehensive treatment of the interaction between a population of
comparable mass embryos will be presented elsewhere (see Zhou et al. 2007).

We choose three sets of models for detail analysis.  In Models 1, 2,
and 3, we set the  mass of each planetesimal to be $m=10^{19}$ g (Model 1),
$10^{24}$ g (Model 2), and $10^{27} $ g(Model 3).  The planetesimals
in Model 2 correspond to the transitional objects as planetesimals
evolve into oligarchies,  which perturb the velocity dispersion of their
neighbors. The embryos in Model 3 correspond to the
isolation mass in a minimum mass nebula interior to the snow line.
 These three representative models
are also chosen to illustrate the relative importance of the gas drag
effects.  The dominant physical process for eccentricity damping is
aerodynamical drag for the low-mass planetesimals in Model 1 and tidal
drag for the embryos in Model 3.  The eccentricity damping time scale
is the longest for the intermediate-mass oligarchies in Model 2.

In order to build up sufficient samples for a statistical analysis, we
use 1000 planetesimals for each simulation. We normalize these models
with $\Sigma_{\rm d} = f_{\rm d} \Sigma_{\rm d, min}$, so the total mass of
planetesimals in the region $[a_{\rm in},a_{\rm out}]$ is:
  \beq
  \bea{ll}M_{\rm tot} & =\int_{a_{\rm in}}^{a_{\rm out}} f_{\rm d} ~\Sigma_{\rm d, min}
  2\pi a da \\
  & =14 f_{\rm d} M_{\oplus}
 \left[ \left(\frac{a_{\rm out}}{\rm ~1AU} \right)^{1/2}
 -\left(\frac{a_{\rm in}}{\rm 1~AU} \right)^{1/2} \right].
 \eea
  \eeq
Thus the total mass of the planetesimals in $[3.2,8.5]$ AU with $f_{\rm d}
=2$ is $31.5~ M_{\oplus}$.

Due to their interaction with each other and the turbulent gas, the
 velocity dispersion of  planetesimals, $\delta {\bf V}= {\bf V}-{\bf V}_{\rm k}$,
is expected to have a Gaussian distribution in Cartesian coordinates.
  The corresponding eccentricities and
inclinations of the test particles follow a Rayleigh distribution
\cite{gl92}. The initial average eccentricities and inclinations of
the planetesimals are taken as $0.0007$ and $0.00035$, respectively.

\section{Formation of planetesimal gap during gas accretion}

For our numerical simulations, we use the Regularized Mixed Variables
Symplectic (RMVS3) method in the SWIFT packet (Levison \& Duncan 1994). %\cite{ld94}.
 This
algorithm is well suited for the computation of close encounters
of planetesimals with  protoplanets (but not the inter-particle encounters).
 The time step of the
integration is set as $0.05$ yr, i.e., $\sim 1/200$ of the period of
the protoplanet orbit.  All planetesimals which venture within the
physical radius of the protoplanet are assumed be
accreted by it.  The density of planetesimals is taken as $\rho_{\rm
tp}= {\rm 2 ~g ~cm^{-3}}$ in equation (\ref{faofin}).  The radius of the
protoplanet is given by equation (\ref{S}]) with two limiting values of $\rho_{\rm p}=
{\rm 1 ~g ~cm^{-3}}$ and $0.125 ~{\rm g ~cm^{-3}}$.  The quantity
$\rho_{\rm p}= {\rm 1 ~g ~cm^{-3}}$ corresponds to that of the core with  radius $R_c$.
 In reality, the effective capture cross section of a protoplanet  is determined
by the planetesimal masses,  the relative speed as well as the protoplanet mass
(which determines the density distribution in its gaseous
envelope). During the initial gas accretion stage of a protoplanet,
its envelope has a relatively small mass and does not significantly
modify the capture cross section.  After the onset of efficient gas
accretion, the density in the outer regions of the envelope is also low as
it undergoes dynamical collapse. Gas accumulates near the core as its
inward motion is halted.  The radius of this location is determined by
the efficiency of radiation transfer but is a few times larger than
$R_c$.  After the protoplanet has attained its asymptotic mass and gas
accretion onto it is quenched, the radius of a typical gas giant
quickly reduces to twice that of Jupiter.  In order to take these
possibilities into account and still keep our investigation within
relatively general and simple bounds, we consider a second set of
capture criterion for intruding planetesimals by doubling the
protoplanet's effective radius (which corresponds to assuming a
homogeneous density $\rho_{\rm p}= {\rm 0.125 ~g ~cm^{-3}}$ for the
protoplanet).

There are also planetesimals scattered to solar distances larger than
$100$ AU or smaller than $1$ AU.  These particles are respectively
classified as ejectors to the outer solar system or Sun crashers.  The
planet mergers, ejectors, and Sun crashers are registered and removed
from the subsequent evolution.

The eccentricity of the planetesimal orbits are excited by the
protoplanet and damped by gas drag.  Eccentricity damping also lead to
a decline in the semi-major axes (see \S\ref{sec:gasdrag}). The energy of
the planetesimal orbits may also be modified by resonant interaction
and close encounters with the protoplanet.  Planetesimals with $a<1$
AU are not able to collide with the protoplanet in their subsequent
evolutions, therefore an inner boundary of $1$ AU is adequate for the
objectives of the present investigation.  In all the simulations
presented here, we assume the gas disk has an initial surface density
of the minimum solar nebula (see Eq. [\ref{rhogas}]). After the formation
of the protoplanet core, a uniform exponential depletion of gas on a
time scale of $10^6$ yr is also assumed.  In most but not all
models, this time scale is larger than the magnitude of $\tau_{\rm
grow}$. The solid disk has an initial surface density of twice the
minimum solar nebula (see Eq. [\ref{dust}]).

\subsection{Opening of planetesimal gap}
\label{sec:pgap}

An important phenomenon in the evolution of planetesimals in a gaseous
environment is the opening of a planetesimal gap around the
protoplanet. The opening of a planetesimal gap in a gas-free environment
was discussed by Rafikov (2001). This situation is analogous to
planetary rings being shepherded by a satellite.  Due to the
differential Keplerian motion, inter-particle encounters lead to
angular momentum transfer and the diffusion of the ring.  But in the
proximity of the protoplanet, tidal perturbation from the protoplanet tends to expel
the planetesimals away from it (Goldreich \& Tremaine 1978, 1980, Lin
\& Papaloizou 1979). When the tidal torque exceeds the rate of
 inter-particle angular momentum exchange, a gap centered on the protoplanet
forms.

During the growth of the embryos and the formation of the core, the
residual planetesimals attain a MRN size distribution, i.e.
most of the planetesimals' surface area and mass are contributed by
the small and large planetesimals respectively.  During the stages of
quasi-hydrostatic sedimentation (S2) and  runaway gas-accretion (S3),
 the maximum embryo size has already increased,  so that the
collision frequency of the smaller residual planetesimals becomes
small compared with that of their synodic encounters with the
protoplanet.  In this low-collision-frequency limit, the
secular perturbation of protoplanet on the residual planetesimals
needs to be taken
into account (Franklin et al. 1980, see \S\ref{sec:feeding}).

Although the energy dissipation rate associated with planetesimal
collisions is reduced with the collision frequency, the excited
planetesimals also experience eccentricity damping from the disk gas.
Since $\tau_{a,e}$ and $\tau_{t, e}$ in equations  (\ref{daaero})
and (\ref{datide}) are much larger than
$\tau_{\rm syno}$ (over which span the eccentricities of
planetesimals are excited to $< \Delta e >$),
the gas drag does not directly reduce the planetesimal eccentricities from
that in equation (\ref{de}).
 However, this process is accompanied
by a slow radially inward drift on time scales $\tau_{a,a} \sim e^{-2}
\tau_{a,e}$ and $\tau_{t,a} \sim e^{-2} \tau_{t,e} $
 (see Eqs. [\ref{daaero}] and [\ref{datide}]) .
  This drift speed is faster for planetesimals
inside than those outside the feeding zone of the protoplanet because they
are excited to larger $< \Delta e >$.

The inward drift causes planetesimals interior to the protoplanet
orbit to drift away from the feeding zone of the protoplanet.  But it also
brings the planetesimals exterior to the protoplanet orbit into its
feeding zone, where their orbital responses become chaotic, resulting
in capture or close encounters.  Planetesimals on both sides of the
protoplanet orbit are evacuated and a gap eventually appears. Beyond
the feeding zone, the planetesimal orbits evolve slowly because the
protoplanet perturbation is much weaker so that the magnitude of $<
\Delta e >$ is much smaller. Some external planetesimals are trapped
onto mean motion resonances of the protoplanet,  where their
eccentricities are increased. Since the gas gap is confined to within
the feeding zone, the residual disk gas damps the planetesimal
eccentricities to modest equilibrium values.

\begin{figure}
\vspace{5.5cm}
%\plotone{f3.eps}
\includegraphics{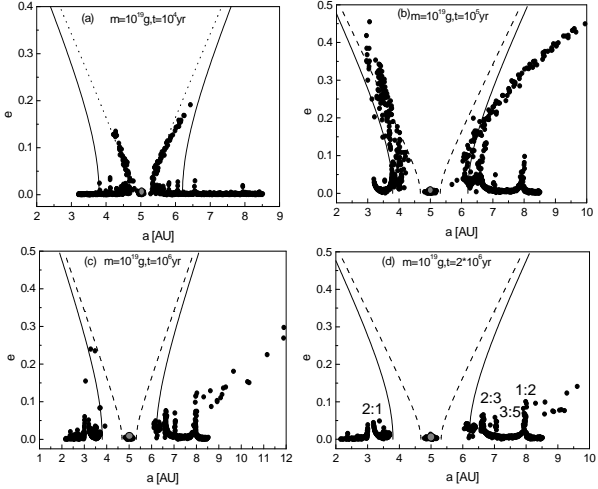}
\caption{\small  Configurations  of survived  planetesimals with  mass
$m=10^{19}$ g in the $(a,e)$ plane during  evolution.
 The protoplanet accretes gas according to the Bondi model on a time scale
 $\tau_{\rm grow}=10^5$ yr.  The
dotted and solid lines are the boundaries of the
  feeding zone ($E=0$) at time $t=10^4$ yr and $t=10^5$ yr,
respectively.
 The gray dot at $a \approx 5$ AU shows the position of the
protoplanet. Some locations of mean motion resonances between planetesimals and the protoplanet
are shown in plot (d).
 \label{fig3}}
\end{figure}

Figure 3 illustrates the distribution of low-mass ($m=10^{19}$ g)
planetesimals (Model 1) in the $(a,e)$ plane at different epochs.  In
this model, we adopt the Bondi gas-accretion prescription and set the
time scale of $\tau_{\rm grow}=10^5$ yr.  At $t=10^4$ yr, $M_{\rm p}
\sim 6 M_\oplus$, $\tau_{\rm syno} \sim 10^3$ yr, and the eccentricities of
the planetesimals within the current feeding zone are excited to
values $< \Delta e > \sim 0.01-0.1$ (Fig.3a).  After $t=10^5$ yr,
the protoplanet acquires its full asymptotic mass.  There are many
planetesimals inside the final feeding zone of the protoplanet. A
planetesimal gap begins to form, albeit at a slower rate than the
expansion of the hypothetical feeding zone (Fig.3b).  At the outer
regions of the feeding zone, the eccentricities of many planetesimals
are greatly excited.  The V shape of the planetesimals' ($a,e$)
distribution indicates that many are scattered analogous to the
KBO's.  After $t \ge 10^6$ yr, planetesimals in the feeding zone
are completely cleared,  as those interior to the protoplanet orbit
drift inward while those exterior to it are scattered outward (Fig.3c,
3d).  The analysis for the clearing time scale of the planetesimal
gap is presented in \S\ref{sec:pgaptime}.

Since the opening of the planetesimal gap is the result of gas drag and the
protoplanet perturbation, the time scale for their eccentricity
damping and therefore gap-clearing depends on their mass (Ida \& Lin
1996).  Figure 4 displays the configurations of survived planetesimals
in the $(a,e)$ plane for planetesimals with mass $m=10^{23}$ g (Model
2) and protoplanetary embryos with $m=10^{27}$ g (Model 3),
respectively.

\begin{figure}
\vspace{5.5cm}
\includegraphics{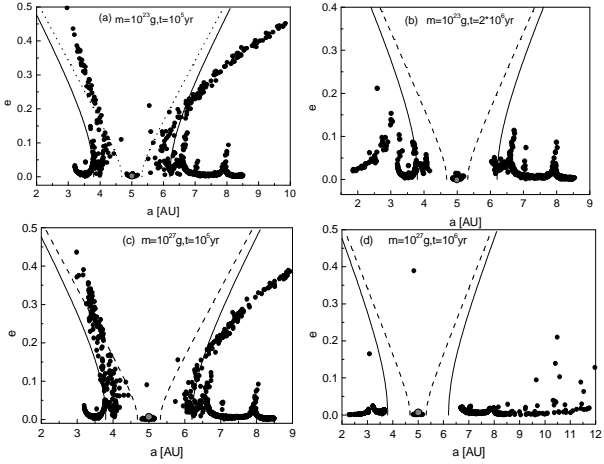}
\caption{\small
Configurations  of  survived planetesimals with  mass
$m=10^{23}$ g and $m=10^{27}$ g in the $(a,e)$ plane during evolution.
 The protoplanet accretes gas by the Bondi model on a time scale
 $\tau_{\rm grow}=10^5$ yr. The
dotted lines and solid lines are the boundaries of the  feeding zone ($E=0$)
 at time $t=10^4$ yr and $t=10^5$ yr,
respectively.  The cases of $t= 10^4 $ yr  are similar to Fig.3.
The gray dot at $a=5$ AU shows the position of the protoplanet.
 \label{fig4}}
\end{figure}

During the buildup of the protoplanet asymptotic mass (at $t=10^5 $
yr in Figs. 4a and 4c), residual planetesimals are found inside the
feeding zone in all models.  A closer inspection indicates that the
opening of the planetesimal gap is less efficient in Model 2 than
in Models 1 and 3.  This minor difference supports the conjecture that
the clearing of the gap is due to the combined action of the protoplanet
perturbation and gas drag,  because the $e$-damping and $a$-decay rates
of the intermediate-mass planetesimals (represented by Model 2) are
smaller than those of the low-mass planetesimals and high-mass embryos
(see Fig.1 for the drift time scale).  On time scales much longer than
$\tau_{\rm grow}$, the intermediate-mass continues to occupy the edge
regions of the feeding zone.  Many planetesimals are also trapped in
the outer mean motion resonances with increased eccentricity
(Fig.4b). In contrast, the drift speed of the embryos ($m=10^{27}$ g)
is faster than those in Models 1 and 2, and all the embryos in the
feeding zone are cleared out after $t=10^6$ yr (Fig.4d).
Nevertheless, the width of the planetesimal gap is limited to $\sim 2
\sqrt{3} h a_{\rm p}$.

\begin{figure}
\vspace{5cm}
\includegraphics{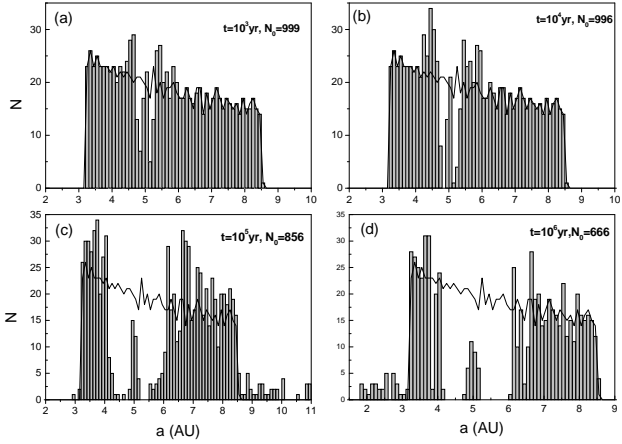}
%\plotone{f5.eps}
\caption{\small
Semi-major axes distribution of survived planetesimals
at four epochs during the mass growth of the protoplanet.
Each bar corresponds to $0.1$ AU. The mass of individual
planetesimal is $10^{23}$ g. The protoplanet accretes gas by the Bondi
model on a time scale
 $\tau_{\rm grow}=10^5$ yr.  $N_0$ is the total number of
planetesimals survived at that time. The solid curve corresponds to the
initial density profile.  The opening of a planetesimal gap leads to a
slight enhancement of surface density near the boundary of the feeding
zone.
 \label{fig5}}
\end{figure}

The opening of a planetesimal gap near the protoplanet is also shown in
Fig. 5 with the intermediate-mass ($10^{23}$ g) planetesimals.  At
$t=\tau_{grow} = 10^5$ yr, the radial distribution of the
planetesimals show a diffusion profile around the corotation radius of
the protoplanet.  Some survival planetesimals near the protoplanet are
 caught onto horseshoe or tadpole orbits.  Figure 6 plots two examples in the tadpole
orbits librating around the $L_4$ and $L_5$ points, respectively.
  In Fig. 5, there is a slight
enhancement of surface density beyond the edge of the feeding zone due
to the resonant trapping of inward-drifting planetesimals, which
will be discussed in \S 4.3. This accumulation of
planetesimals increases the local surface density and $M_{\rm iso}$,
 promotes the growth rate of protoplanetary cores and the
emergence of secondary proto-gas-giant planets.

\begin{figure}
 \vspace{5cm}
 \includegraphics{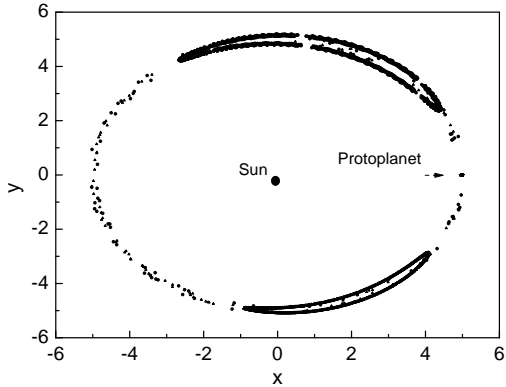}
 %\plotone{f6.eps}
 \caption{\small Tadpole
orbits of two surviving planetesimals in model Fig.5d.  The
coordinate is set as origin at the mass center of the sun and protoplanet (with mass growing),
 and corotating  with the protoplanet at
$\approx 5AU$.  The planetesimals' asymptotic eccentricity is
$e\approx 0.001 $.  The dot around $(5,0)$ marks the
orbit of the protoplanet.  The planetesimals have semi-major axis
$a=4.875$ and $a=5.038$ at $t=10^6$ yr, respectively.
%$e=1.9\times 10^{-3}, i=0.90\times 10^{-3}0.00090$, and
%$a=4.805$, $e=5.2\times 10^{-3}, i=1.6\times 10^{-3}0.00090$ at time $t=10^6$ yr.
 \label{fig6}}
\end{figure}

\subsection{Planetesimal gap clearing time scale}
\label{sec:pgaptime}

There are several relevant time scales to be considered.  In
\S\ref{sec:feeding}, we showed that, under the protoplanet
perturbation, planetesimals with overlapping 2-body (neglecting the
stellar tide) orbits extend to the boundary of the inner feeding zone
where $\delta_a < 1.5 h$.  The time scale of depletion of the total
planetesimal population in this region, $M_{\rm p, f} =4 \pi \Sigma_{\rm d}
\delta_a a_{\rm p}^2$, can be derived  in view of  equation (\ref{eq:mdotp}) with
$e \sim <\delta e> $  from  equation (\ref{de}),
 \beq
 \bea{ll}
 \tau_{\rm p, f} & \equiv  {M_{\rm p, f} \over {\dot M}_{\rm p} }
  \simeq \left( {3 P_k \over 4 \pi} \right) \left( {\delta_a \over h}
 \right)^3  \cdot \\
   & \left( 1 + {4 \over 9} \left( { 1.9 h \over \delta_a} \right)^6
 \right) \left( {a_{\rm p} \over R_c} \right)
 \simeq \left( {a_{\rm p} \over R_c} \right) P_k
 \eea
 \eeq
where $P_k$ is the Keplerian period. So $\tau_{\rm p, f} < 10^5$ yr.

This rapid depletion time scale is only applicable to
the planetesimals in the inner feeding zone where $\delta_a < 1.5 h$.
Planetesimals in the outer feeding zone with $\delta_a/h \sim 1.5
-3.5$ only occasionally venture into the Roche
lobe of the protoplanet. Based on the discussions in the previous section, we now derive
the time scale for planetesimals to migrate across the boundary
between the inner and outer feeding zone.  Since $\tau_{\rm p, f}$ is
relatively small, the duration of the migration across the outer
feeding zone essentially corresponds to the clearing
time scale of the planetesimal gap.

Suppose the protoplanet has a mass $\mu=M_{\rm p}/M_\ast$ at time $t$,
so  its instantaneous normalized Hill radius is $h=(\mu/3)^{1/3}$.
In terms of the protoplanet's asymptotic (at $t\ge \tau_{\rm grow}$)
normalized Hill's radius $h_{\rm f}$, the scaled distance of a planetesimal
is defined as
 \beq
 b_{\rm f}\equiv \frac{\delta_a}{h_{\rm f}},
 \label{bf}
 \eeq
where $\delta_a=|a/a_p-1|$. According to equation (\ref{daaero}),
 the speed of the inward drift for a
planetesimal with mass $m\le 10^{23}$ g is given as
 \beq
 \dot{a}_{\rm aero}\approx - 1.7 \frac{a_{\rm p} }{\tau_{\rm
 aero}} e ^3,
 \label{dadt}
 \eeq
where we suppose $e^2 \gg \eta$, because the eccentricity of
planetesimals inside the feeding zone could be excited up to $\sim
0.1$ (e.g., Figs.3-4).  In \S\ref{sec:feeding}, we evaluate the average
excursions of $<\Delta e>$ per encounter. (As small initial inclination
is expected, we neglect the modulation in $\Delta i$.) Since $\tau_{a,e}
> > \tau_{\rm syno}$, we can replace $e$ in equation (\ref{dadt}) by $<\Delta e>$
in equation  (\ref{de}) to obtain
 \beq
 (\dot{b}_{\rm f})_{\rm aero}
  = \frac{550 }{\tau_{\rm aero}}\frac{1}
  { b_{\rm f}^6 }\frac{h^9}{h_{\rm f}^7},
  \eeq
where the dependence on $h_{\rm f}$ is introduced for the purpose of
normalization.  With the growth of the protoplanet mass, $h$ also
increases with time. However, in order to simplify the problem, we
first assume $h$ as a constant $h_{\rm f}$ (an approximation to be justified
{\it a posteri}) so that
 \beq
 \left(b_{\rm f} \right)_{\rm aero}=\beta_1 \left(\frac{h}{h_{\rm f}}
 \right)   \left(\frac{h^2 }
 {\tau_{\rm aero}}t \right)^{1/7},  ~~(b_{\rm f} \le 2\sqrt{3} \left(\frac{h}{h_{\rm f}} \right) ),
 \label{dbaein}
 \eeq
where $\beta_1=3.2$.  The limiting value ($2 \sqrt{3}$) for $b_{\rm f}$
corresponds to the width of the entire feeding zone. Planetesimals
drift from the protoplanet orbit to the inner boundary of the
asymptotic feeding zone and from the outer boundary of the asymptotic
feeding zone to the protoplanet orbit on a time scale of
 \beq
 \left(T_{\rm open} \right)_{\rm aero}=
(\frac{2\sqrt{3}}{\beta_1})^7\frac{\tau_{\rm aero}}{h^2}.
\label{topenae}
 \eeq
The above expression is obtained by equating the left hand size of equation
(\ref{dbaein}) with $2\sqrt{3} h/h_{\rm f}$.

 According to the above
equation, gap formation for the low-mass planetesimals proceeds on a
time scale $\tau_{aero}/h^2$. Substituting $\tau_{\rm aero}$ from equation
(\ref{tgas}) and assuming a constant protoplanet  mass, we find
$T_{\rm open} \sim 3 (M_J/M_{\rm p})^{2/3}$ Myr for Model 1.  We carry out
two comparisons, Models 1a and 3a, in which the mass of the protoplanet
is fixed to that of Jupiter so that $h=h_{\rm f}$. We compare, in Figs. 7a
and 7c, the results of the numerical simulations with that in
equation (\ref{dbaein}). Numerically, we determine $b_{\rm f}$ from the
distribution plots such as Fig. 5 at some typical epoch.  The
magnitude of $b_{\rm f}$ is defined to be the maximum half-width of the gap
within which only planetesimals on horseshoe or tadpole orbits (around
$5$AU) survived.

The qualitative agreement between the functional dependence of $b_{\rm f}$ on $h$ in
the numerical results and the expression in equation  (\ref{dbaein}) supports
the interpretation that the clearing of the gap is regulated by the
orbital decay of the planetesimals in the feeding zone. (This
agreement is particularly good during the expansion of the gap through
the initial outer feeding zone where the migration scenario is most
applicable.)  However,  in comparison with the expression in equation
(\ref{dbaein}), a factor of $2-3$ for $\beta_1$
is needed to fit the numerical simulations. This difference in the
magnitude of $\beta_1$ is caused by an underestimate in the analytical
expression for $<\Delta e>$ which did not take into account the
cumulative eccentricity modulation during each secular cycle.
From equation ({\ref{topenae}) , the expansion of the gap is
stalled with an asymptotic width $\sim 2 \sqrt{3} h$ at $\sim
\tau_{\rm aero}$ in the numerical simulations.  A similar analysis also
applies to the massive embryos for which the tidal drag is more
appropriate.

We now return to the more realistic models in which the planetesimal
gap formation proceeds during the growth of the protoplanet. Despite
the increases of the protoplanet mass, the above approximations
would essentially be valid if $\tau_{\rm grow} > T_{\rm open}$.  In
this limit, planetesimal gaps with the instantaneous feeding zone
width ($2\sqrt{3} h$) form and expand with the mass of the
protoplanet.  Based on the assumption that most planetesimals in the
feeding zone can collide with the protoplanet core and are the main
contributors to the initial growth of the protoplanet, Pollack et al.
(1996) derived their bombardment rate onto the core from the
expansion rate of its feeding zone. In that model, the
suppression of the gas accretion rate due to the energy dissipation of
planetesimal accretion has been taken into account.

However, in the limit of modest $M_{\rm p}$ (a few $M_\oplus$), both
the protoplanet's $h$ and the planetesimal eccentricities are very small so that
$T_{\rm open} > \tau_{\rm grow}$ even for a protracted protoplanetary
growth.  In this case, the feeding zone expands faster than they can
be cleared out (especially in Model 2 in Fig.4).  Both the protoplanet
mass and feeding zone width attain their asymptotic values on a time
scale $\tau_{\rm grow}$, after which gap formation and clearing of the
planetesimals in the feeding zone proceed on a time scale $T_{\rm
open}$.  This gradual mass ramp up significantly delays the clearing
of the gap.  During the advanced stage, the assumption of constant
$M_{\rm p}$ is again satisfied and the values of $h$ can be replaced by
$h_{\rm f}$ in the above equations.  In the next section, we show that not
all the planetesimals in the feeding zone collide with the
protoplanet core and their collision rate may be substantially lower
than that estimated by Pollack et al. (1996). This effect can
reduce the barrier for the gas accretion rate onto the protoplanet
envelope.

Using the same approach, we deduce the width and time scale associated
with the gap-opening of intermediate and high-mass ($>10^{23}$ g)
embryos.  In this case, the gravitational tidal drag provides the
dominant eccentricity damping effect. From equation (\ref{datide}),  we find
 \beq
  \dot{a}_{\rm tidal}\approx - \frac{5}{8}
  \frac{a_{\rm p}  }{\tau_{\rm tidal }} e ^2,
  \label{datd}
 \eeq
and
\beq
  \left(b_{f} \right)_{\rm tidal}= \beta_2  \left(\frac{h}{h_{\rm f}} \right)
\left(\frac{h }{\tau_{\rm tidal}}t \right)^{1/5}, ~~(b_{\rm f} \le  2\sqrt{3} \left(\frac{h}{h_{\rm f}} \right)),
  \label{dbtdin}
\eeq where $\beta_2=2.7$. The time scale to open a gap is given as,
 \beq
 \left(T_{\rm open} \right)_{\rm tidal}=(\frac{2\sqrt{3}}{\beta_2})^5 \frac{\tau_{\rm tidal}}{h}.
 \eeq
 Figure 7 shows the evolution of the width of the planetesimal gap
determined from our numerical simulations of Models 1 and 3.  In
Figs. 7c-7d, the protoplanet accretes gas according to the Bondi
accretion prescription with $\tau_{\rm grow}=10^5$ yr, as in the
cases of Figs. 3-5. The solid curves are theoretical predictions
given by equations (\ref{dbaein}) and (\ref{dbtdin})  with different
coefficients.  The functional forms again are in general agreement,
though there is a factor of 3 discrepancy in the coefficient of
$\beta_2$.

\begin{figure}
\vspace{5cm}
\includegraphics{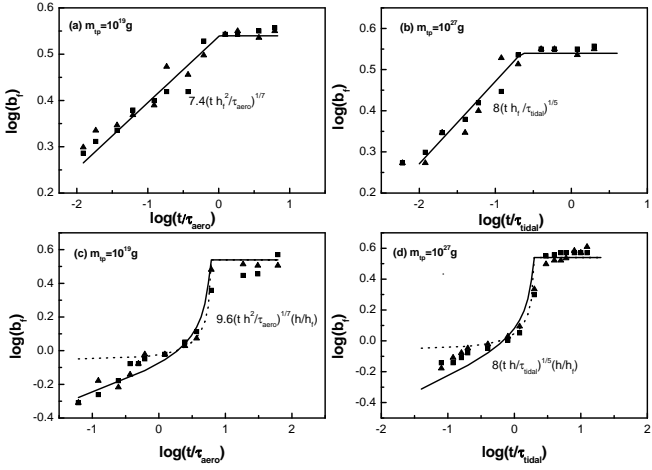}
%\plotone{f7.eps}
\caption{\small Evolution of the width of the planetesimal gap.
The protoplanet has constant mass $\mu=10^{-3}$ in panels (a) and (b), and grows from
$\mu=1.7\times 10^{-5}$ to $\mu=10^{-3}$ in panels (c) and (d) through gas-accretion
 with a Bondi gas-accretion model of $\tau_{\rm grow}=10^5$.   The solid curves are theoretical
predictions given by equations (\ref{dbaein}) and (\ref{dbtdin}) with
coefficients: (a) $\beta_1=7.4$, (b) $\beta_2=8.0$, (c) $\beta_1=9.6$,
(d) $\beta_2=8.0$.
The squares and triangles   denote the inner and outer boundaries, respectively.
The dashed curves in panels (c) and (d) show the width of the
feeding zone defined by $2\sqrt{3}|a-a_{\rm p}|/(a_{\rm p}h)$ at time $t$.
\label{fig7}} \end{figure}

Outside the boundary of the feeding zone, the rate of the eccentricity
excitation by the protoplanet is small. According to equations
(\ref{daaero}) and (\ref{datide}), the migration speed becomes
correspondingly smaller by several orders of magnitude. Though the
migration speed outside the gap can be obtained following similar
lines of reasoning, such an analysis is not crucial for this study and
we will not discuss it further.

\subsection{Dependence on gas accretion model and time scale }

In the above simulations, the time scale of gas accretion onto the
protoplanet, $\tau_{\rm grow}$, is set to be $10^5$ yr.  In order
to determine the dependence of the collision efficiency of
planetesimals on the growing time scale of the protoplanet, we adopt a
range of $\tau_{\rm grow}$, from $10^3$ to $10^6$ yr, with both the
Bondi (\ref{bd}) and linear (\ref{ln}) prescriptions for gas
accretion.  Figure 8 illustrates the collision ($P_{\rm col}$) and escape
($P_{\rm esc}$) probabilities of intermediate-mass ($10^{23}$ g)
planetesimals (in Model 2) during the growth of the protoplanet.  Since the
evaluation of the protoplanet radius is based on the present
 density of gas giants,  $P_{\rm col}$ should be regarded as a lower limit.
Although protoplanets have extended envelopes, most of their mass
resides in the core and the density in the envelope decreases rapidly
with radius.  Small and modest-size planetesimals may be captured by
the protoplanet when they pass through its envelope.  But the large
embryos can only merge with the protoplanet through direct collisions
with the core.

\begin{figure}
\vspace{5cm}
\includegraphics{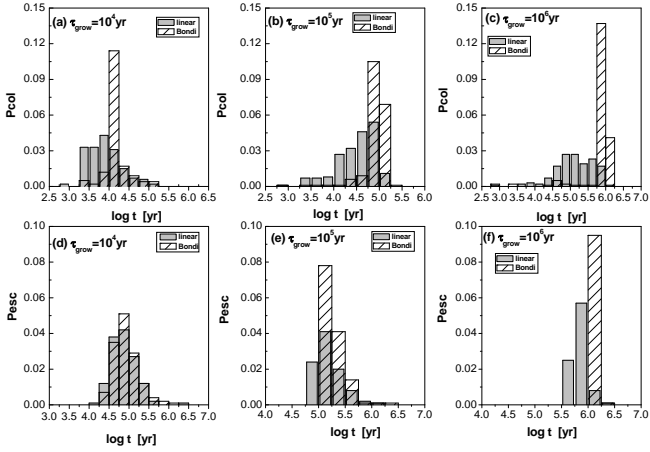}
\caption{\small Distributions of the planetesimal collision probability
$P_{\rm col}$ (a-c) and escape probability $P_{\rm esc}$ (d-f) as
a function of the evolution time. The time scales for
gas accretion  are $\tau_{\rm grow}=10^4,10^5,10^6$ yr, respectively.
In the Bondi gas-accretion model, the planetesimals collide with
the protoplanet mainly around the late stage of gas accretion, which
is quite different from that in the linear gas-accretion
model.  \label{fig8}}
\end{figure}

Throughout the evolution, $T_{\rm open} > \tau_{\rm grow}$,  so that the
expansion of the feeding zone outpaces the clearing process (Figs. 3-4).
Consequently, major epochs for planetesimal collisions with the
protoplanet occur around $t=\tau_{\rm grow}$ in Bondi gas-accretion
model (Fig.8).  The duration of the epoch of intense bombardment is
$\sim 10^{\log (\tau_{\rm grow}) \pm 0.25}$ for $\tau_{\rm grow} \ge
10^{5}$ yr.  In contrast, major collision  events would occur
much earlier if the protoplanet mass increases according to the
hypothetical linear gas-accretion prescription.  Most of these
collisions occur at $ t \le \tau_{\rm grow}$. The dichotomy between
these two types of accretion arises because the Bondi
prescription leads to a runaway process,  in which most of
the protoplanet  mass is attained only at the very end when
$t=\tau_{\rm grow}$. However, the hypothetical linear accretion
introduces a much earlier ramp up in the
protoplanets mass. Consequently its physical radius, gravitational
perturbation, and width of feeding zone also grow quickly, inducing
an earlier phase of intensified collisions.

Also note that both $P_{\rm esc}$ and $P_{\rm col}$ are normalized to the
initial planetesimal population in the entire computational domain
which covers twice the width of the asymptotic feeding zone of the
protoplanet. The total cumulative statistics suggest that,
 the fraction of the
original planetesimal population in the feeding zone which collides
with the protoplanet is comparable to that scattered to the outer disk
regions.  With both prescriptions, the ejections of planetesimals
occur around $t \ge \tau_{\rm grow}$.  For a given density, the
 radius and surface escape speed $V_{\rm esp}$ of the
 protoplanet are
proportional to $M_{\rm p}^{1/3}$.  During the initial growth stages of the
protoplanet, its $V_{\rm esp}$ is small compared with the local
Keplerian velocity.  Scattering from grazing encounters excite the
planetesimal eccentricities rather than eject them.  At an advanced
growth stage of the protoplanet(when its $M_{\rm p} \sim M_J$), however,
scattering with impact parameter larger than a few planetary radii
can lead to large-angle deflections and the escape of planetesimals
(Lin \& Ida 1997).  The ratio of $P_{\rm esc}/P_{\rm col}$ would increase with
 the semi-major axis ($a_{\rm p}$) and effective density ($\rho_{\rm p}$)
 of the protoplanet(Ida \& Lin 2004).
The latter quantity is likely to increase after the gas accretion onto
the protoplanet envelope is quenched by its tidal barrier
(Dobbs-Dixon et al. 2007).

\begin{figure}
\vspace{5cm}
\includegraphics{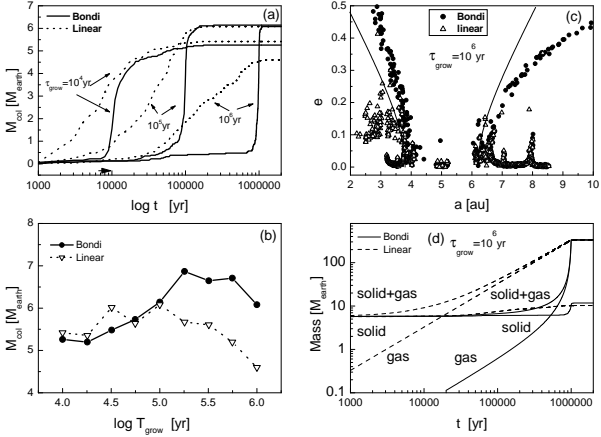}
  \caption{\small Dependence of
 collided planetesimal mass on $\tau_{\rm grow}$ and gas accretion model.
  (a) Evolution of the planetesimal
mass colliding with the protoplanet during its gas accretion.
 (b) Variation of collided solid mass with the
time scale of gas accretion $\tau_{\rm grow}$ in the two different
models.  (c) A comparison of the survived planetesimals with mass
$m=10^{23}$ g in Bondi and linear gas-accretion models during the
evolution.  (d) Evolution of solid and gaseous compositions of the
protoplanet. In this paper, we suppose the solid and gaseous disks
have a surface density two times and one time the minimum solar
nebula, respectively.
 \label{fig9}}
\end{figure}

We now scale our model with the minimum mass nebula model. By adopting
the same radial dependence but twice the magnitude in surface density
as that in the minimum mass nebula, the total solid mass in the region
$a\in [3.2,8.5]$ AU  is $31.5 ~M_{\oplus}$ in
our models(\S 2.3).  We deduce the rate of accretion by multiplying this total
solid mass with the planetesimal accretion probability.  Figure 9a
shows the solid mass ($M_{\rm col}$) that collides with the protoplanet
as a function of its growth time scale ($\tau_{\rm grow}$).  With
Bondi accretion, the magnitude of $M_{\rm col}$ attains
a maximum value with $ \tau \in [10^5, 10^6]$ yr (Fig. 9b), which is
$6 \sim 7 ~M_{\oplus}$. (Once again, these values are
applied to compact protoplanets and should be regarded as lower
limits.)  With the same $\tau_{\rm grow}$, the collided solid mass in
the linear accretion prescription is $1\sim 2 ~M_{\oplus}$ less
because the more rapid initial growth of the protoplanet leads to
early excitation and evacuation of the feeding zone. This rapid
initial ramp up of the protoplanet mass causes many planetesimals to
drift inward due to gas drag or be ejected before the feeding zone
attains its asymptotic width (Fig.9c).  In the limit of large
$\tau_{\rm grow}$, a protoplanet with Bondi gas-accretion model can
accrete planetesimals more effectively.

Figure 9d shows the evolution of solid (planetesimals) and gas
 (including dust grains) component accreted by the protoplanet.  The
protoplanet has more solid mass (including the initial core which is
$\sim 6 M_\oplus$) in the beginning of Bondi accretion due to the
inefficient gas accretion. But after $M_{\rm p}=M_{\rm sod} + M_{\rm gas} > 7
M_\oplus$, the gas accretion rate far exceeds that of the planetesimal
accretion rate because $T_{\rm open} > \tau_{\rm grow}$.  While the accreted
gas may carry small dust particles with it, the suppression of initial
planetesimal accretion reduces the feedback effect which limits the
gas accretion rate. This late addition
of planetesimals can lead to the enrichment of protoplanet envelope
to super-solar metallicity.

\begin{figure}
\vspace{5cm}
\includegraphics{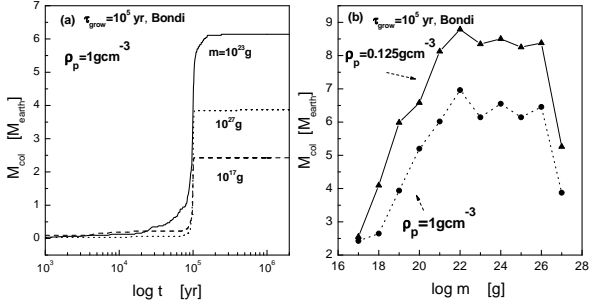}
\caption{\small  Dependence of  collision rate on  planetesimal mass ($m$).
 (a) Evolution of the total solid mass collided onto
 the protoplanet  with individual planetesimal mass
ranging from $10^{17}-10^{27}$ g in the Bondi gas-accretion model of
$\tau_{\rm grow}= 10^5$ yr. (b) Dependence of the collided
planetesimals' mass on their individual masses.
 \label{fig10}}
\end{figure}

We now examine the dependence of planetesimal accretion efficiency on
their mass $m$. This dependence arises because the eccentricity
damping time scale is a function of $m$ (see Fig.1). Figure 10 shows
the evolution of collided solid mass with individual planetesimal mass
ranging from $10^{17}\sim 10^{27}$ g. The maximum collision rate occurs
with individual planetesimal mass $m\in 10^{20\sim 26}$ g,
which corresponds to planetesimals with radius of $50\sim 5000 $
km. These planetesimals also have the largest $\tau_{\rm aero}$ and
$\tau_{\rm tidal}$.  The planet with smaller density ($\rho =0.125$ g cm$^{-3}$) has a bigger collision rate.
 In Fig. 11, we plot the
probabilities of the planetesimals survived, collided, ejected or
crashed after $2\times 10^6$ yr in a Bondi accretion model with
$t_{\rm grow}=10^5 $ yr. We find that at least $~2/3$ of
planetesimals have survived at an epoch $\sim 10 t_{\rm
grow}$.  In this model, the initial width of the planetesimal disk is
$15 a_{\rm p} h_f$ out of which a planetesimal gap with a width $\sim 7 a_{\rm p} h_f$
is evacuated. Although the final distribution of the planetesimal disk
is more extended (see Fig.5), most of the surviving planetesimals $(>2/3)$
are located within a range of nearly $ 8 a_{\rm p} h_f$ from the protoplanet.
Thus, the density outside the feeding zone is slightly enhanced on
average after the protoplanet obtains its asymptotic mass.

\begin{figure}
\vspace{5cm}
\includegraphics{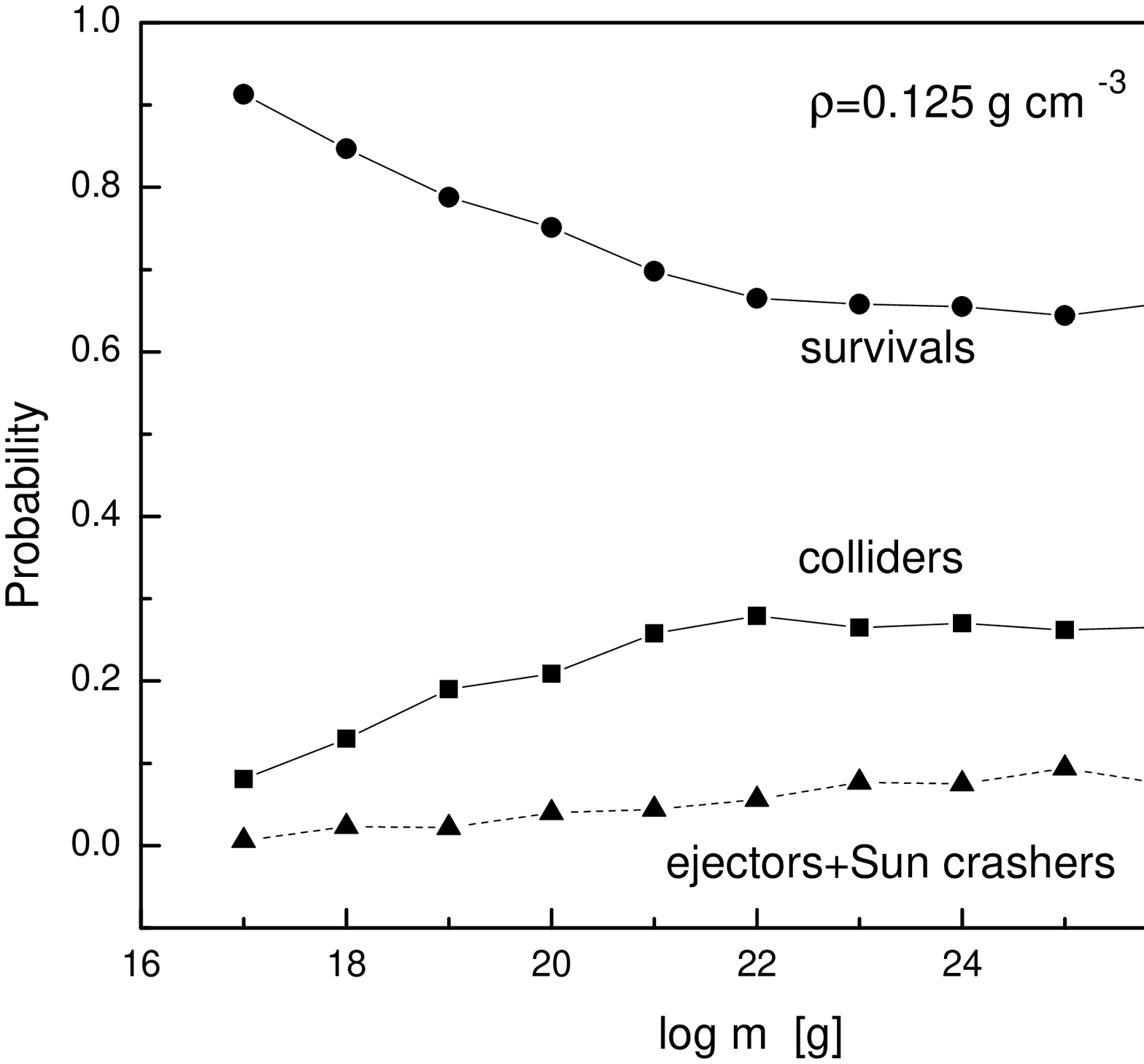}
 \caption{\small
Probability of survivers, colliders, ejectors($a>100$ AU) and
 Sun cashers ($a<1$ AU) as functions of the planetesimals' mass $m$ in the range
$10^{17}-10^{27}$ g at $t=2\times 10^6 $yr.
The gas-accretion model is Bondi accretion. The model parameters are
$\tau_{\rm grow}= 10^5$ yr, protoplanet density $\rho=0.125$ g cm$^{-3}$.
 \label{fig11}}
\end{figure}

\subsection{Resonant trapping}
\label{sec:resonance}

In both Figures 3 and 4, the accumulation of planetesimals near the
outer mean motion resonances of protoplanet is noticeable, particularly
for the intermediate-mass Model 2. The first-order mean motion
resonances are located both interior and exterior to $a_{\rm p}$ where the
period ratio can be expressed either as $(p+1):p$ or $p:(p+1)$.
The corresponding ratio of semi-major axes with $a_{\rm p}$ is $\alpha =
(1+1/p)^{-2/3} $ or $(1+1/p)^{2/3}$ respectively.  In each
case, the protoplanet mass grows from 5.6$M_\oplus$ to 300 $M_\oplus$. With its
asymptotic mass, the 2:1 and 3:2 resonances of the protoplanet  are outside
its feeding zone,  whereas mean motion resonances up to 10:9 are beyond
$2 \sqrt{3} h$ at the onset of the simulations.

Using a simple pendulum model (\S8, Murray \& Dermott 1999), the width
and libration time scale associated with these mean motion resonances
can be derived as
 \beq
 \Delta_{\rm res} = \left( {16 R_r (\alpha) \mu
 e_{\rm res} \over 3} \right)^{1/2} a_{\rm p} ,
 \eeq
 \beq
 \tau_{\rm res} = \left( {3 p^2 e_{\rm res} R_r (\alpha) \mu }
 \right)^{-1/2} P_{\rm k},
 \eeq
where $P_{\rm k}$ is the Keplerian period, and the magnitude of $R_r(\alpha)=\alpha
[p+1+\frac12 \alpha D] b^{p+1}_{1/2}$ is
an increasing function of $p$.  The minimum value of eccentricity in the
resonance is $< \Delta e>$ (see Eq. [\ref{de}]).
During the resonant passage, an adiabatic invariant constrains
the eccentricity change to be
 \beq
 \Delta e_{\rm res}^2 = {\Delta_{\rm res}  \over p a_{\rm p}} .
 \eeq
In the limit that $ \Delta e_{\rm res}  $ due to resonant
perturbation is larger than $ < \Delta e >$ due to secular
perturbation, it can be substituted by $e_{\rm res}$ so that
 \beq
 e_{\rm res} \simeq \left( {16 R_r (\alpha) \mu \over 3 p^2}
 \right)^{1/3},
 \eeq
 \beq
 \Delta_{\rm res} = \left( {16 R_r (\alpha) \mu \over 3}
 \right)^{2/3} {a_{\rm p} \over p ^{1/3}},
 \label{eq:reswidth}
 \eeq
 \beq
 \tau_{\rm res} = 12^{-1/3} \left( p  R_r (\alpha) \mu
 \right)^{-2/3} P_{\rm k}.
 \eeq

Substituting $e_{\rm res}$ into equations (\ref{daaero}) and (\ref{datide}), we
find that the damping of the resonantly excited eccentricity leads to
a characteristic migration time across $\Delta_{\rm res}$, which is
 \beq
 \tau_x = {\Delta_{\rm res} \over {\dot a_{\rm aero} (e_{\rm res}) }}
 = \left( \frac{ 16 p R_r (\alpha) \mu}{3} \right)^{1/3}\frac{\tau_{aero}}{2}
 \eeq
for small planetesimals. For the more massive embryos,
  \beq
 \tau_x = {\Delta_{\rm res}  \over {\dot a_{\rm tide} (e_{\rm res}) }}
  = { 8\over 5 } p \tau_{\rm tide},
 \eeq
which is independent of the protoplanet mass.

Planetesimals are trapped in resonances when $\tau_x > \tau_{\rm res}$.  As
a planetesimal approaches a protoplanet, it encounters resonances with
increasing $p$ and $R_r(\alpha)$,  so  the magnitude of $\tau_{\rm
res}$ decrease with little change in the magnitude of $e_{\rm
res}$. In principle, it should be easier for the strong resonance
close to the protoplanet to capture the planetesimals because $\tau_x$
becomes larger than $\tau_{\rm res}$ for sufficiently large
$p$'s. However, for relatively large $p$'s, the normalized distance
separating the $p:(p+1)$ and $(p-1):p$ mean motion resonances,
 \beq
{\Delta_{p, p-1} \over a_{\rm p}} \simeq { 2 \over 3 p^{2} },
 \eeq
is a decreasing function of $p^2$ but independent of the protoplanet mass.
In contrast, equation (\ref{eq:reswidth}) indicates that the magnitude of
$\Delta_{\rm res}/a_{\rm p}$ increases with $\mu = M_{\rm p}/M_\ast$
and decreases with $p^{1/3}$.
When the protoplanet attains
 \beq
\mu > \mu_{p, c} = 2^{-5/2} 3^{-1/2} R_r(\alpha)^{-1} p^{-5/2},
 \label{eq:overlap}
 \eeq
the width of its $p:(p+1)$ resonance (i.e. $\Delta_{\rm res}$) becomes
larger than the separation between it and the $(p-1):p$ resonance
(i.e. $\Delta_{p, p-1}$).  Overlapping resonances generally lead
to dynamical instabilities which excite the eccentricities of the
trapped planetesimals and modify their semi-major axes (Murray \& Dermott
1999).

The expression in equation (\ref{eq:overlap}) indicates that the critical mass
for the overlapping resonances is a decreasing function of $p$. During
the growth of the protoplanet,  planetesimals located in the initial
inner feeding zone become destabilized and collide with the protoplanet
first.  But the planetesimals captured onto the more distant low-order
mean motion resonances may remain trapped during the growth of the protoplanet.
 For example, the resonant capture condition is most easily
satisfied for the ``distant'' 3:2 and 2:1 mean motion resonances.  As
the protoplanet grows, the resonance
strengthen increases and the libration time scale is reduced.  Both
the resonant probability and the characteristic eccentricity of the
resonant planetesimals increase.  In the standard model with
$\tau_{\rm grow} = 10^5$ yr, the mass doubling time scale prior to the termination
of growth is $\sim 10^3$ yr,  which is comparable to the libration
time scale in the mean motion resonances.  Planetesimals that are
loosely bound to the mean motion resonances have longer libration time
scale than $\tau_{\rm res}$. They are shaken by the rapid evolution of
the protoplanet's gravitational potential and cannot respond through
adiabatic adjustments.  This impulse leads to a late episode of
planetesimal bombardment,  which can introduce metallicity and structure
diversity to the gas giant planets.  The impact of this impulsive
shake up is most pronounced in the rapid gas accretion models (see Figs.
8a and 8d), where the resonant capture becomes ineffective.

The condition for resonant trapping is also more easily satisfied for
the intermediate mass planetesimals,  because their eccentricity damping
and orbital migration time scales are relatively long.  These
tendencies are clearly evident in Figs. 3 and 4.  During the
epoch of gas depletion, the damping time scales $\tau_{\rm aero}$ and
$\tau_{\rm tide}$ lengthen, which again provide favorable conditions for
the capture of residual planetesimals into the mean motion resonances.

Finally, the excess density in the mean motion resonances is determined
by the migration rate outside the resonances. Through orbital decay,
planetesimals from the regions outside the first-born protoplanet
congregate near its mean motion resonance.  The enhancement of the
local surface density promotes the formation of secondary gas giants.

\begin{figure}
\vspace{5cm}
\includegraphics{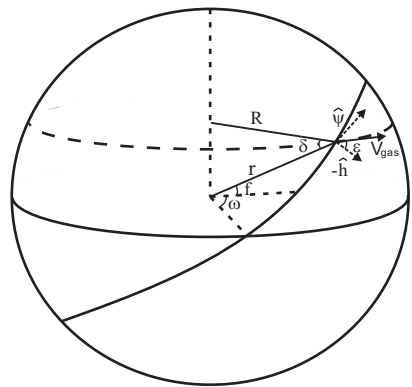}
%\vspace{-7cm}
\caption{\small
Configuration of tidal drag perturbation on a planetesimal (see Appendix).
 \label{fig12}}
\end{figure}

\section{Summary and discussion}

In this paper, we investigate numerically the process and
efficiency of planetesimal accretion onto a growing protoplanet.  We
consider the stage after the formation of the protoplanet core and
during the accretion of its envelope. We use a physical Bondi formula
and an {\it ad hoc} linear prescription to evaluate the gas accretion
rate.  The seed protoplanet is placed at $5$ AU with initial mass
$5.67~M_{\oplus}$, accreting gas in a time scale $\tau_{\rm
grow}$. The results of our numerical simulation and analysis have
implications for three issues:

\subsection{Suppression of planetesimal accretion during the onset of
gas accretion.}

We first examine the initial growth of the protoplanet  through gas
accretion (transition from stages of embryo-growth (S1) to quasi-hydrostatic sedimentation
(S2)).  The accretion rate of the
gas is suppressed by the bombardment of planetesimals,  which generates
heat needed to be redistributed efficiently.  In the vicinity of
the core, there are two regions that supply the bullet planetesimals.
Through its secular perturbation, the protoplanet induces
planetesimals within an inner feeding zone to attain radial excursion,
which crosses its orbit during each azimuthal conjunction.
Planetesimals within this band ($\sim 1.5 h a_{\rm p}$) engage in repeated
close encounters. At 5AU, the local Keplerian speed is comparable to the
 surface escape speed of the protoplanet, so only a fraction of the close
encounters will lead to physical collisions and the buildup of the core.

The protoplanet also has an outer feeding zone at $1.5-3.5 h
a_{\rm p}$.  Due to the tidal perturbation of the host star and the
protoplanet, planetesimals in this region occasionally cross the orbit
of the protoplanet.  However, the frequency of such encounters
decreases with $\delta_a$ and vanishes at the boundary of the feeding
zone.  In a gas free-environment, many synodic periods are needed for
the planetesimals in the outer feeding zone to be captured by the
protoplanet.  The presence of gas can lead to eccentricity damping,
orbital migration, and diffusion of planetesimals from the outer to
the inner feeding zone.  The migration time scale is determined by the
  gravitational perturbation of the  protoplanet as well as the efficiency of
gas damping. At the onset of the gas-accretion stage, the mass of the
protoplanetary core is relatively low,  so the  excited
eccentricities of the planetesimals are relatively small.
Consequently the migration rate from the outer to the inner feeding zone is slow.

The low replenishment rate as well as the modest collision probability imply
that the planetesimal bombardment rate onto the core is much less
frequent than that inferred from the efficient consumption of all
planetesimals engulfed by the expanding feeding zone (as assumed in
the early models of Pollack et al. 1996).  The suppression of
planetesimal collisions also lead to a decline in the energy
dissipation rate and a cutoff in the replenishment of grains in the
gaseous envelope of the protoplanet.  The elimination of these bottlenecks
for gas accretion shortens the growth time scale for proto-gas-giant
planets.

\subsection{Chemical and structural diversity through late-stage
bombardment}

A substantial fraction of the super-solar chemical composition in
Jupiter resides in its envelope.  We explore the possibility that the
chemical and core-envelope structural diversity may be due to
late-stage bombardment by residual planetesimals.  Due to the secular
perturbations of the protoplanet and gas drag, exterior planetesimals
(with $a > a_{\rm p}$) in the outer feeding zone diffuse into the inner
feeding zone and engage in close encounters with the protoplanet.  The
interior planetesimals (with $a < a_{\rm p}$) undergo orbital decay out of
the feeding zone.  Both effects lead to the clearing of the feeding
zone and the formation of a planetesimal gap.

In tenuous regions of the disk, protoplanet grows gradually and the
orbits of both planetesimals and the massive embryos decay slowly.
The migration of the external planetesimals and embryos may be stalled
near the outer mean motion resonances of the protoplanet.  But during
the growth of the protoplanet, its feeding zone expands and the mean
motion resonances (with high $p$'s) overlap.  The
planetesimals accumulated in the resonances become dynamically
unstable. The colliding embryos may penetrate deeply into the
protoplanet envelope and become part of its core.

However, in the relatively dense inner regions of the disk, the
gas-accretion rate of protoplanet and the  orbital decay
rates of the planetesimals are both relatively high.
Embryos pass through the mean motion
resonances without any significant perturbation.  Only the
intermediate-mass planetesimals can be captured onto the mean motion resonances
of the protoplanet. When the protoplanet becomes sufficiently
massive to destabilize the resonance-trapped planetesimals, the planetesimals
will collide with the protoplanet. However, as they do not
have adequate mass to survive the passage through its envelope,
 they mostly supply the metallicity in the
protoplanet envelope.

These two possible outcomes may account for the structural diversity
between Jupiter and Saturn.  Jupiter formed early in a relatively dense
region just beyond the snow line.  Intermediate-mass planetesimal
bombardments occurred during the advanced stage of its progenitor's
growth so that it has a relatively low core mass but a metal-rich
envelope.  In contrast, Saturn formed during the depletion epoch in
the relatively tenuous outer regions of the disk.  In this case,
massive embryos may be trapped in the mean motion resonances
of the protoplanet  when
its mass is a few $M_\oplus$.  Their orbits  become unstable
when Saturn acquired most of its present-day mass.  The late-stage
bombardment by these massive embryos may have contributed to
substantial core of Saturn.

Generally the longer the time scale of gas accretion, the more
efficient a protoplanet accretes planetesimals.  But when the time
scale is comparable to the age of the gaseous disk (millions of years), a
runaway type of gas accretion model like Bondi accretion is preferred
for the protoplanet to acquire more efficient and late-stage
planetesimal accretions (Figs.8-9).  In a solid disk with surface
density twice the minimum solar nebula, the solid mass
colliding with a compact protoplanet is $6\sim 7 ~M_{\oplus}$,
and it reaches maximum when $ \tau \in [10^5, 10^6]$ yr in the Bondi
gas-accretion model.  This mass is comparable to the initial core mass
and should be regarded as a lower limit.
  In our calculations, the radius is calculated
according to equation (\ref{S}).  After the protoplanet acquired an atmosphere, the
collision rate could be enhanced (Inaba \& Ikoma 2003) and a higher
solid accretion rate may be expected.

The accretion rates of planetesimals with different masses are
determined.  The mass of individual planetesimal that could be
effectively accreted to the protoplanet lies in the range
$10^{20}-10^{26}$ g, which corresponds to  an embryo  with
radius of $50-5000 $ km (Fig.10).  The relatively low mass
planetesimals disintegrate in the protoplanet envelope,  whereas
massive protoplanetary embryos may survive their passage through
the envelope and become a part of the protoplanet core.

\subsection{The enhanced formation of multiple gas giant planetary
systems}

We also show that due to the gas drag and protoplanet secular
perturbations, the density of the planetesimal disk could be slightly
increased outside the orbit of the protoplanet, which will enhance the
subsequent formation of external protoplanet cores (Figs.3-5).

During their orbital decay, all planetesimals, with the
exception of the most massive embryos,  migrate sufficiently slowly that
they become locked onto the  mean motion resonances of the protoplanet.  When
the growth of the protoplanet is stalled, the resonant planetesimals residing
outside the asymptotic feeding zone remain well separated from other
mean motion resonances and are stably trapped in these
resonances.  The enhancement of the local surface density reduces the
 growth time scale of the embryos.  The formation of
second generation proto-gas-giant planets  is promoted.

The emergence of multiple gas giants close to each other's mean motion
resonances may also lead to dynamical instabilities (Zhou et al. 2007).
The resulting dynamical interaction may lead to mergers,
ejections, and breakup of the system.  We suggest that this may be a
promising avenue for the generation of the large eccentricity
distribution among the extra solar planets.

\acknowledgments

We thank Drs. M. Nagasawa, S. Aarseth and S. Dong for constructive
discussions, and Dr. S. Aarseth for improving the manuscript.
 This work is supported by NSFC(10233020, 10778603), NCET (04-0468),
NASA (NAGS5-11779, NNG04G-191G, NNG06-GH45G), JPL (1270927), NSF(AST-0507424, PHY99-0794).

\appendix

\section{Perturbations under tidal drag}

 Suppose the perturbing  acceleration of tidal  drag to a planetesimal
  has the form of equation (\ref{tidal}):
 \beq
f_{\rm tidal}=-\frac{{\bf V}_k-{\bf V}_{\rm g}}{\tau_{\rm tidal}}
\equiv A({\bf V}_k-{\bf V}_{\rm g}),
\label{tidal2}
\eeq
where $A= 1/\tau_{\rm tidal} $. The velocity of the planetesimal and gas can be expressed in terms of the
radial, azimuthal and normal components with unit vectors $\hat{r},\hat{\psi}$  and $\hat{h}$,
respectively (Fig. 12):
\beq \begin{array}{l}
{\bf V}_k=v_0 (\cos \alpha ~\hat{r} +\sin \alpha  ~\hat{\psi} ), \\
{\bf V}_{\rm g}=\chi (\cos \epsilon  ~\hat{\psi} +\sin \epsilon ~\hat{h} ),
\end{array}
\eeq
where $\alpha$ is the angle  from the radial to the velocity direction in the planetesimal
orbit plane, and  $\chi=\sqrt{GM_{\odot}/R}(1-2\eta (R))^{1/2}$. Suppose
the gas is in a circular orbit,
$\eta=0$, we obtain,
\beq \chi=n a (\frac{1+e\cos f}{1-e^2})^{1/2} \cos ^{-1/2} \delta.
\eeq
The definitions of the angles are  also shown in Fig.12.
From spherical geometry (Adachi et al. 1976),
\beq
\begin{array}{l}
 \sin \epsilon = \cos i /\cos \delta , \\
 \cos \epsilon = \sin i \cos (f+\omega) / \cos \delta, \\
 \cos \delta = [1-\sin ^2 i \sin ^2(f+\omega )]^{1/2},
\end{array}
\eeq
where $f$ is the true anomaly of the planetesimal orbit.
Thus the perturbing acceleration of tidal drag can be expressed as:
\beq
\begin{array}{rl}
f_{\rm tidal}= & \frac{naA}{\sqrt{1-e^2}}\{ e\sin f \hat{r}+(1+e\sin f)^{1/2}[(1+e\sin f)^{1/2}
-\cos i \cos ^{-3/2}\delta  ] \hat{\psi}  \\
& +(1+e\sin f)^{1/2} \sin i \cos (f+\omega) \cos ^{-3/2}\delta \hat{h}\} \\
\equiv &  \bar{R} \hat{r} +\bar{T} \hat{\psi} +\bar{N} \hat{h}
\end{array}
\eeq
The  evolution equations of
the osculating elements under tidal drag obey (Murray \& Dermott 1999):
\begin{equation}
\begin{array}{ll}
\frac{da}{dt} & = \frac{2}{n\sqrt{1-e^2}} [ \bar{R} e\sin f +\bar{T} (1+e\sin f)] \\
&= \frac{2aA}{1-e^2}\{ 1+e^2+2e\cos f-(1+e\sin f)^{3/2}\cos i
[1- \sin^2 i \sin^2 (f+\omega)]^{-3/4} \}\\
\frac{de}{dt} &= \frac{\sqrt{1-e^2}}{na} [\bar{R}\sin f + \bar{T} (\cos f+\frac{\cos f+e}{1+e\cos f})] \\
\ & =2A(e+\cos f) +A(1+e\cos f)^{-3/2}(2\cos f + e\cos^2 f+e)
\cos i [1- \sin^2 i \sin (f+\omega)]^{-3/4} \\
\frac{di}{dt}& = \frac{\sqrt{1-e^2}}{na}\frac{\cos (f+ \omega)}{1+e\cos f}\bar{N} \\
 &= A\sin i (1+e\cos f)^{-1/2} [1- \sin^2 i \sin (f+\omega)]^{-3/4} \cos^2 (f+\Omega).
\end{array}
\eeq
We average over time $t$ in one period of Keperian motion according to :
\beq
<F>=\frac{1}{T}\int_{0}^{T}F dt =\frac{1}{2\pi}\int_{0}^{2\pi}\frac{F(1-e^2)^{3/2}}{(1-e\cos f)^{2}} df.
\eeq
This gives:
\beq
\begin{array}{ll}
\frac{1}{a}<\frac{da}{dt}> & = \frac{A}{8}[(5e^2+2 i^2) +o(e^2,i^2)],  \\
\frac{1}{e}<\frac{de}{dt}> & = A [(1-\frac{13}{32} e^2-\frac12 i^2+\frac34 i^2 \sin^2\omega)+o(e^2,i^2)],  \\
\frac{1}{i}<\frac{di}{dt}> & = \frac{A}{2}[ (1-\frac{13}{32} e^2+\frac{35}{16}e^2\sin^2\omega
+\frac{3}{16}i^2)+o(e^2,i^2)]. \\
\end{array}
\eeq
We further average over one period of $d\omega/dt$ to eliminate the dependence of $\omega$,
 and recall $A=-1/\tau_{\rm tidal}$, which finally yields
\beq
\begin{array}{ll}
\frac{1}{a}<\frac{da}{dt}> & = -\frac{1}{8\tau_{\rm tidal}}[(5e^2+2 i^2)+o(e^2,i^2)],   \\
\frac{1}{e}<\frac{de}{dt}> & = -\frac{1}{\tau_{\rm tidal}} [(1-\frac{13}{32} e^2-\frac{1}{8}i^2)+o(e^2,i^2)],  \\
\frac{1}{i}<\frac{di}{dt}> & = -\frac{1}{2\tau_{\rm tidal}}[ (1+\frac{11}{16} e^2+\frac{3}{16}i^2)+o(e^2,i^2)] . \\
\end{array}
\eeq


\begin{thebibliography}{}
%\bibitem[Aarseth, Lin \& Palmer 1993]{ALP93} Aarseth, S.J.,  Lin  D.N.C., \& Palmer, P.L. 1993, ApJ, 403,351
\bibitem[Adachi, Hayashi \&  Nakazawa 1976]{ahn76} Adachi, I., Hayashi, C., \& Nakazawa, K. 1976,
Prog. Theo. Phys. 56, 1756
\bibitem[Artymowicz(1993)]{art93}Artymowicz, P., 1993, \apj, 419,155
%\bibitem[Bodenheimer 1974]{bod74}Bodenheimer, P. 1974, Icarus, 23, 319
%\bibitem[Bodenheimer 1980]{bod80}Bodenheimer, P. 1980, Protostars and Planets II.
% eds. D. C. Black \& M. S. Matthews (Tucson : Univ. Arizona Press), 873
\bibitem[Bodenheimer, Lin, \&  Mardling 2001]{blm01}
 Bodenheimer, P., Lin, D. N. C., \&  Mardling, R. A. 2001,\apj, 548, 466
\bibitem[Bodenheimer \& Pollack 1986]{bp86}Bodenheimer, P., \& Pollack, J. B. 1986, Icarus, 67, 391
\bibitem[Burrows et al. 2000]{bur00}Burrows, A., et al. 2000, \apj, 534, L97
% Burrows, A.; Guillot, T.; Hubbard, W. B.; Marley, M. S.; Saumon, D.; Lunine, J. I.; Sudarsky, D.
\bibitem[Bryden et al. 2000]{bry00} Bryden, G., et al. 2000, \apj, 540, 1091
%\bibitem[Bryden, Lin \& Ida 2000]{bli00}Bryden, G., Lin, D.N.C. ,\&  Ida, S. 2000, ApJ, 544, 481
%\bibitem[Cameron 1978]{cam78}Cameron, A.G.W. 1978, Moon Planets, 18, 5
%\bibitem[DeCampli \& Cameron 1979]{dc79}DeCampli, W.M. and Cameron, A.G. 1979, Icarus 38, 367
\bibitem[Dobbs-Dixon, Li \&  Lin 2007]{dll07}Dobbs-Dixon, I.,  Li, S. L., \& Lin, D. N. C. 2007,
Tidal Barrier and the Asymptotic Mass of Proto Gas-Giant Planets, arXiv:astro-ph/0701269
\bibitem[Frank, King \& Raine 2002]{fkr02} Frank, J., King, A., \& Raine, D.  2002,
 Accretion power in Astrophysics. (Cambridge: Cambridge Univ. Press)
\bibitem[Franklin {\it et al.} 1980]{fra80}
Franklin, F. A., et al.  1980, Icarus, 42, 271 %Lecar, M.,  Lin, D. N. C., Papaloizou, J.
\bibitem[Garaud \&  Lin  2007]{gl07}Garaud, P., \&  Lin, D. N. C. 2007, \apj, 654, 606
 \bibitem[Goldreich, Lithwick \& Sari 2004 ]{gls04}
 Goldreich, P.,  Lithwick, Y., \&  Sari, R. 2004, \apj,  614, 497
  \bibitem[Goldreich\ & Tremaine 1978]{gt78}Goldreich, P., \& Tremaine, S.  1978,
  \apj, 222, 850
 \bibitem[Goldreich\ & Tremaine 1980]{gt80}Goldreich P., \& Tremaine, S.  1980,
 \apj, 241, 425
 %Goldreich, Peter; Lithwick, Yoram; Sari, Re'em
\bibitem[Greenzweig \& Lissauer 1992]{gl92}Greenzweig, Y., \& Lissauer, J. J. 1992, Icarus, 100, 440
\bibitem[Guillot, Gautier \& Hubbard 1997]{ggh97}Guillot, T., Gautier,D., \& Hubbard, W. B. 1997, Icarus, 130, 534
\bibitem[Hasegawa \& Nakazawa 1990]{hn90}Hasegawa, M., \& Nakazawa, K. 1990,  A\&A, 227, 619
 \bibitem[Hayashi, Nakazawa \& Nakagawa 1985]{hay85}Hayashi, C., Nakazawa, K. \&  Nakagawa, Y. 1985,
  in {\rm Protoplanets
 and Planets II.} ed. D. C. Black \& M. S. Mathew (Tucson: Univ. Arizona Press), 1100
\bibitem[Hayashi 1981]{hay81}Hayashi, C. 1981, Prog. Theo. Phys. Suppl. 70, 35
\bibitem[Hayashi, Nakazawa \& Adachi 1977]{kna77} Hayashi, C., Nakazawa K. \& Adachi, I. 1977,
Publ. Astron. Soc. Japan, 29, 163
%\bibitem[Hayashi, Nakazawa \& Nakagawa 1985]{hnn85} Hayashi, C., Nakazawa K.,\& Nakagawa, Y. 1985,
%Protostars and Planets II. eds. D. C. Black \& M. S. Matthews (Tucson : Univ. Arizona Press), 1100
\bibitem[Hubickyj, Bodenheimer \& Lissauer, 2005]{hub05}Hubickyj, O., Bodenheimer, P., \& Lissauer, J. J. 2005,
Icarus, 179, 415
 %RevMexAA, 22, 83-86
\bibitem[Ida \& Lin 1996]{il96} Ida, S., \& Lin, D. N. C. 1996, AJ, 112, 1239
\bibitem[Ida \& Lin 2004]{il04} Ida, S., \& Lin, D. N. C. 2004, \apj, 604, 388
\bibitem[Ida \& Lin 2007]{il07} Ida, S., \& Lin, D. N. C. 2007, \apj, submitted
%Toward a Deterministic
%Model of Planetary Formation IV: Core's Migration,
%Self-Regulated Clearing and the late Epoch of Gas Giant Formation,
\bibitem[Ida \& Makino 1993]{im93}Ida, S., \& Makino, J. Icarus,  106,  210
\bibitem[Ikoma, Nakazawa \&  Kiyoshi 2000]{ink00}
 Ikoma, M., Nakazawa, K., \& Emori, H. 2000, \apj,  537, 1013
 %Ikoma, Masahiro; Nakazawa, Kiyoshi; Emori, Hiroyuki
\bibitem[Inaba \& Ikoma 2003] {ii03}Inaba, S., \& Ikoma, M. 2003, A\&A, 410, 711
\bibitem[Kokubo \& Ida 1996]{ki96} Kokubo, E., \& Ida, S. 1996, Icarus, 123, 180
\bibitem[Kokubo \& Ida 1998]{ki98} Kokubo, E., \& Ida, S. 1998, Icarus, 131, 171
\bibitem[Kokubo \& Ida 2000]{ki00} Kokubo, E., \& Ida, S. 2000, Icarus, 143, 15
\bibitem[Kokubo \& Ida 2002]{ki02} Kokubo, E., \& Ida, S. 2002, \apj, 581, 666
\bibitem[Koller, Li \& Lin 2003]{kll03}Koller, J., Li, H., \& Lin, D. N. C  2003, \apj, 596, L91
\bibitem[Kominami \& Ida 2002]{koi02}Kominami, J., \& Ida, S. 2002, Icarus 157,43
%\bibitem[Kuiper 1951]{kui51} Kuiper G. P., 1951, In {\em Astrophysics: A Topical Symposium }(J.A.Hynek, Ed.) pp.357-324.
%  McGraw-Hill, New York
%\bibitem[Landau \& Lifthitz 1999]{l999}Landau,L.D., \& Lifthitz, E.M.  1999, Fluid Mechanics, Butterworth-Heinemann
\bibitem[Laughlin, Steinacker \& Adams 2004]{lsa04}
Laughlin, G., Steinacker A., \&  Adams, R. 2004, ApJ, 608, 489
\bibitem[Levison \& Duncan 1994]{ld94}Levison, H. F., \& Duncan, M. J. 1994, Icarus 108, 18
\bibitem[Lin , Bodenheimer \& Richardson (1996)]{lbr96}
Lin, D. N. C, Bodenheimer, P., \& Richardson, D. C. 1996, Nature, 380, 606
\bibitem[Lin \& Ida 1997]{li97} Lin, D. N. C., \& Ida, S. 1997 \apj, 477, 781
 \bibitem[Lin \& Papaloizou  1979]{pl79}
 Lin, D. N. C, \& Papaloizou, J. C. B. 1979, MNRAS, 186, 799
\bibitem[Lin \& Papaloizou 1986]{lp86}
 Lin, D. N. C, \& Papaloizou, J. C. B. 1986, \apj, 309, 846
\bibitem[Lin \& Papaloizou 1993]{lp93}
 Lin, D. N. C, \& Papaloizou, J. C. B. 1993, in
\rm{Protostars and Planets III}, eds. E. H. Levy  \& J.I. Lunine,
 (Tucson: Unv. Arizona Press),  p.\ 749
\bibitem[Leinhardt \& Richardson 2005]{lr05}
Leinhardt, Z., \&  Richardson, D. C. 2005, \apj, 625, 427
\bibitem[Lissauer 1987]{lis87}Lissauer, J. J. 1987, Icarus, 69, 249
\bibitem[Marcy et al. 2005]{mar05}Marcy, G., et al. 2005,
 Prog. Theo. Phys. Suppl.  158, 24
\bibitem[Mizuno 1980]{Miz80}Mizuno, H. 1980, Prog. Theo. Phys.  64, 544
\bibitem[Murray \& Dermott 1999]{md99}Murray, C. D., \& Dermott, S. F. 1999, Solar System Dynamcis (Cambridge: Cambridge Univ. Press)
\bibitem[Nelson \& Papaliozou]{np03} Nelson, R. P., \& Papaloizou, J. C. B. 2003, MNRAS,  339, 993
\bibitem[Palmer, Aarseth \& Lin 1993]{pal93} Palmer, P. L.,  Lin, D. N. C., \&  Aarseth, S. J. 1993,
      \apj,  403, 336
\bibitem[Perri \&Cameron 1974]{PC74} Perri, F., \& Cameron, A. G. W. 1974, Icarus, 22, 416
\bibitem[Pollack {\it  et al.} 1996]{pol96} Pollack, J. B., et al. 1996, Icarus, 124, 62
\bibitem[Rafikov 2001]{raf01}Rafikov, R. R. 2001, AJ, 122, 2713
\bibitem[Rice \& Armitage 2003]{ra03} Rice, W. K. M., \& Armitage, P. J. 2003, \apj, 598, L55
\bibitem[Safronov 1969]{saf69}Safronov, V.S. 1969, Evolution of the Protoplanetary Cloud and Formation of
   the Earth and Planets (Moscow: Nauka)
\bibitem[Sato et al. 2005]{sato05}Sato, B., et al. 2005, \apj, 633, 465
\bibitem[Saumon \& Guillot 2004]{sg04}Saumon, D., \& Guillot, T. 2004, ApJ, 609, 1170
\bibitem[Shakura \& Sunyaev 1973]{ss73}Shakura, N. I., \& Sunyaev, R.A. 1973, AAp, 24, 337
%\bibitem[Thommes, Duncan \& Levison 2003]{tdl03} Thommes, E.W., Duncan, M.J., \& Levison, H.F. 2003, Icarus 161, 431
%\bibitem[V$\ddot{o}$lk et al. 1980 ]{volk80}V$\ddot{o}$lk, H.J., Jones, F.C., Morfill, G.E., \& R$\ddot{o}$ser, S. 1980,
%A\&A, 85,316
\bibitem[Tanaka \& Ida 1999]{ti99}Tanaka, H., \& Ida, S. 1999, Icarus, 139, 350
\bibitem[Tanaka,  Takeuchi \& Ward 2002]{ttw02}Tanaka, H., Takeuchi, T., \& Ward, W. R.
 2002, \apj, 565, 1257  %Tanaka, Hidekazu; Takeuchi, Taku; Ward, William R.
\bibitem[Ward 1986]{wd86}Ward, W. R. 1986, Icarus, 67, 164
\bibitem[Ward 1989]{wd89}Ward, W. R. 1989, \apj, 336, 526
\bibitem[Ward 1997]{wd97}Ward, W. R. 1997, Icarus, 126, 261
%\bibitem[Weidenschilling 1984]{Wei84} Weidenschilling, S. 1984, Icarus, 60, 553
%\bibitem[Weidenschilling, \& Cuzzi 1992]{WC92} Weidenschilling, S. \& Cuzzi, J.N. 1992, Protostars and Planets III. eds. E. Levy  \& M. Matthews (Tucson : Univ. Arizona Press), 1031
\bibitem[Wetherill \& Stewart 1989]{ws89}Wetherill, G. W., \& Stewart, G. R. 1989, Icarus, 77, 330
\bibitem[Wuchterl, Guillot \& Lissauer, 2000]{wgl00} Wuchterl, G., Guillot, T., \& Lissauer,J. J. 2000,
in Protostars and Planets IV eds. V. Mannings, A. P. Boss \& S. S. Russell (Tucson : Univ. Arizona Press), 1081
\bibitem[Zhou, Lin \& Sun 2007]{zls07}Zhou, J. -L., Lin, D. N. C., \& Sun Y. -S. 2007,
% Post-Oligarchic Evolution of  Protoplanetary Embryos and the Stability of
%      Planetary Systems,
      \apj, accepted (astro-ph 0705.2164)

\end{thebibliography}
\end{document}